\documentclass[useAMS,usenatbib]{mnras}

\usepackage{ae,aecompl} 
\usepackage{epsfig}
\usepackage{multirow}
\usepackage{multicol}
\usepackage{longtable}
\usepackage{threeparttablex}
\usepackage{lscape}
\usepackage{amssymb,amsmath}
\usepackage{url}
\usepackage[usenames]{color}
\usepackage[]{hyperref}

\newcommand{\bd}{\begin{displaymath}}
\newcommand{\ed}{\end{displaymath}}
\newcommand{\be}{\begin{equation}}
\newcommand{\ee}{\end{equation}}
\newcommand{\beaa}{\begin{eqnarray*}}
\newcommand{\eeaa}{\end{eqnarray*}}
\newcommand{\bea}{\begin{eqnarray}}
\newcommand{\eea}{\end{eqnarray}}

\def\hequad{HE\,0435$-$1223}

\def\Om{\Omega_{\rm m}}
\def\OL{\Omega_{\Lambda}}

\def\tdist{D_{\Delta t}}

\def\Dd{D_{\rm d}}
\def\Dds{D_{\rm ds}}
\def\Ds{D_{\rm s}}

\def\kext{\kappa_{\rm ext}}
\def\gext{\gamma_{\rm ext}}

\def\zd{z_{\rm d}}
\def\zs{z_{\rm s}}

\newcommand{\vect}[1]{\boldsymbol{#1}}

\newcommand{\chid}{\chi_{\mathrm{d}}}
\newcommand{\chis}{\chi_{\mathrm{s}}}

\newcommand{\cadd}{f_{\mathrm{d}}}
\newcommand{\cads}{f_{\mathrm{s}}}
\newcommand{\cadds}{f_{\mathrm{ds}}}

\newcommand{\cadK}{f_{K}}

\newcommand{\vx}{\vect{x}}
\newcommand{\vtheta}{\vect{\theta}}

\newcommand{\deltam}{\delta_{\mathrm{m}}}
\newcommand{\deltag}{\delta_{\mathrm{g}}}
\newcommand{\sigmag}{\sigma_{\mathrm{g}}}
\newcommand{\biasg}{b_{\mathrm{g}}}

\title[The mass along the line of sight to the gravitational lens \hequad]{H0LiCOW III. Quantifying the effect of mass along the line of sight to the gravitational lens \hequad\ through weighted galaxy counts
\thanks{Based on data collected at Subaru Telescope, which is operated by the National Astronomical Observatory of Japan. }}

\author[C.E.~Rusu et al.]
{Cristian E. Rusu,$^{1}$\thanks{E-mail: cerusu@ucdavis.edu}
Christopher D. Fassnacht,$^{1}$   
Dominique Sluse,$^{2}$
Stefan Hilbert,$^{3,4}$
 \newauthor
Kenneth C. Wong,$^{5,6}$
Kuang-Han Huang,$^{1}$
Sherry H. Suyu,$^{6,7}$
Thomas E. Collett$^{8}$
 \newauthor
Philip J. Marshall,$^{9}$ Tommaso Treu$^{10}$ and Leon V. E. Koopmans$^{11}$\\ 
$^1$Department of Physics, University of California, Davis, CA 95616, USA\\ 
$^{2}$STAR Institute, Quartier Agora - All\'ee du six Aout, 19c B-4000 Li\'ege, Belgium\\
$^{3}$ Exzellenzcluster Universe, Boltzmannstr. 2, 85748 Garching, Germany\\
$^{4}$ Ludwig-Maximilians-Universit{\"a}t, Universit{\"a}ts-Sternwarte, Scheinerstr. 1, 81679 M{\"u}nchen, Germany\\
$^{5}$National Astronomical Observatory of Japan, 2-21-1 Osawa, Mitaka, Tokyo 181-8588, Japan\\
$^{6}$Institute of Astronomy and Astrophysics, Academia Sinica (ASIAA), P.O.~Box 23-141, Taipei 10617, Taiwan\\
$^{7}$Max-Planck-Institut f{\"u}r Astrophysik, Karl-Schwarzschild-Str.~1, 85748 Garching, Germany\\
$^{8}$Institute of Cosmology and Gravitation, University of Portsmouth, Burnaby Rd, Portsmouth PO1 3FX, UK\\
$^{9}$Kavli Institute for Particle Astrophysics and Cosmology, Stanford University, 452 Lomita Mall, Stanford, CA 94035, USA\\ 
$^{10}$Department of Physics and Astronomy, University of California, Los Angeles, CA 90095-1547, USA\\
$^{11}$Kapteyn Astronomical Institute, University of Groningen, PO Box 800, NL-9700 AV Groningen, The Netherlands\\
}

\date{Accepted XXX. Received YYY; in original form ZZZ}
\pubyear{2016}

\begin{document}
\pagerange{\pageref{firstpage}--\pageref{lastpage}} 
\maketitle
\label{firstpage}

\begin{abstract}
Based on spectroscopy and multiband wide-field observations of the gravitationally lensed quasar \hequad , we determine the probability distribution function of the external convergence $\kext$ for this system. We measure the under/overdensity of the line of sight towards the lens system and compare it to the average line of sight throughout the universe, determined by using the CFHTLenS as a control field.  Aiming to constrain $\kext$ as tightly as possible, we determine under/overdensities using various combinations of relevant informative weighing schemes for the galaxy counts, such as projected distance to the lens, redshift, and stellar mass. We then convert the measured under/overdensities into a $\kext$ distribution, using ray-tracing through the Millennium Simulation.
We explore several limiting magnitudes and apertures, and account for systematic and statistical uncertainties relevant to the quality of the observational data, which we further test through simulations. Our most robust estimate of $\kext$ has a median value $\kappa^\mathrm{med}_\mathrm{ext} = 0.004$ and a standard deviation of $\sigma_\kappa = 0.025$. The measured $\sigma_\kappa$ corresponds to $2.5\%$ uncertainty on the time delay distance, and hence the Hubble constant $H_0$ inference from this system. The median $\kappa^\mathrm{med}_\mathrm{ext}$ value is robust to $\sim0.005$ (i.e. $\sim0.5\%$ on $H_0$) regardless of the adopted aperture radius, limiting magnitude and weighting scheme, as long as the latter incorporates galaxy number counts, the projected distance to the main lens, and a prior on the external shear obtained from mass modelling.    
The availability of a well-constrained $\kext$ makes \hequad\ a valuable system for measuring cosmological parameters using strong gravitational lens time delays. 
\end{abstract}



\section{Introduction}\label{section:intro}

By measuring time delays between the multiple images of a source with time-varying luminosity, strong gravitational lens systems with measured time delays can be used to measure cosmological distances and the Hubble constant $H_0$ \citep{Refsdal64}.  In particular, for a lens system with a strong deflector at a single redshift, one may infer the \lq{}time-delay distance\rq{}
\begin{equation}
\tdist = (1+\zd) \frac{\Dd \Ds}{\Dds},
\label{eq:tdist}   
\end{equation}
where $\zd$ denotes the redshift of the foreground deflector, $\Dd$ the angular diameter distance to the deflector, $\Ds$ the angular diameter distance to the source, and $\Dds$ the angular diameter distance between the deflector and the source. The time-delay distance is primarily sensitive to the Hubble constant, i.e. $\tdist \propto H_0^{-1}$ \citep[see][for a recent review]{treu16}.

Inferring cosmological distances from measured time delays also requires accurate models for the mass distribution of the main deflector and its environment, as well as for any other matter structures along the line of sight that may influence the observed images and time delays \citep{suyu10}.
Galaxies very close in projection to the main deflector often cause measurable higher-order perturbations in the lensed images and time delays and require explicit models of their matter distribution. The effect of galaxies more distant in projection is primarily a small additional uniform focusing of the light from the source. Furthermore, matter underdensities along the line of sight such as voids, indicated by a low galaxy number density, cause a slight defocusing.
For a strong lensing system with a main deflector at a single redshift, the net effect of the (de)focusing by these weak perturbers is equivalent (to lowest relevant order) to that of a constant external convergence\footnote{The external convergence $\kext$ may be positive or negative depending on whether focusing or defocusing outweighs the other.} term $\kext$ in the lens model for the main deflector \citep{suyu10}. This implies on the one hand that the weak perturbers' effects, i.e. the external convergence they induce, cannot be inferred from the observed strongly lensed image properties alone due to the \lq{}mass-sheet degeneracy\rq{} \citep[MSD,][]{falco85,schneider13}. 
On the other hand, if the external convergence is somehow determined from ancillary data, and a time-delay distance $\tdist^{(0)}$ has been inferred using a model not accounting for the effects of weak perturbers along the line of sight, the true time-delay distance $\tdist$ can simply be computed by:
\begin{equation}
\tdist = \frac{\tdist^{(0)}}{1 - \kext}.
\label{eq:tdist_and_external_convergence}   
\end{equation}
This relation makes clear that any statistical and systematic uncertainties in the external convergence due to structures along the line of sight directly translate into statistical and systematic errors in the inferred time delay distance and Hubble constant:
\begin{equation}
H_0 = (1 - \kext) H_0^{(0)},
\label{eq:H}   
\end{equation}
where $H_0^{(0)}$ denotes the Hubble constant inferred when neglecting weak external perturbers. With reduced uncertainties on other component of the time delay distance measurement from state-of-the-art imaging, time-delay measurements, and modeling techniques of strong lens systems, the external convergence $\kext$ is now left as an important source of uncertainty on the inferred $H_0$, contributing up to $\sim 5\%$ to the error budget on $H_0$ \citep{suyu10,suyu13}. Moreover, the mean external convergence may not vanish for an ensemble of lens systems due to selection effects, causing a slight preference for lens systems with overdense lines of sights \citep{collett16}. Thus, an ensemble analysis simply assuming $\kext=0$ is expected to systematically overestimate the Hubble constant $H_0$. 

Accurately quantifying the distribution of mass along the line of sight requires wide-field imaging and spectroscopy \citep[e.g.,][see \citet{treu16} for a recent review]{keeton04,fassnacht06,momcheva06,fassnacht11,wong11}. \citet{suyu10} pioneered the idea of estimating a probability distribution function $P(\kext)$ by (i) measuring the galaxy number counts around a lens system, (ii) comparing the resulting counts against those of a control field to obtain relative counts, and (iii) selecting lines of sight of similar relative counts, along with their associated convergence values, from a numerical simulation of cosmic structure evolution.
To this end, \citet{fassnacht11} measured the galaxy number counts in a $45\arcsec$ aperture around \hequad\ [$\alpha$(2000):~04h~38m~14.9s, $\delta$(2000):~-12$^\circ$17\arcmin14\farcs4; \citealp{wisotzki00,wisotzki02}; lens redshift $\zd = 0.455$; \citealp{morgan05}; source redshift $\zs = 1.693$; \citealp{sluse12}], and found that it is 0.89 of that on an average line of sight through their $\sim0.06 \deg^2$ control field. Both \citet[][hereafter G13]{greene13} and \citet{collett13} find that $P(\kext)$ can be most precisely constrained for lens systems along underdense lines of sight, making \hequad\ a valuable system. 

Recent work has focused on tightening the constraints on $P(\kext)$ with data beyond simple galaxy counts. \citet{suyu13} used the external shear inferred from lens modelling as a further constraint, which significantly affected the inferred external convergence due to the large external shear required by the lens model. G13 extended the number counts technique by considering more informative, physically relevant weights, such as galaxy redshift, stellar mass, and projected separation from the line of sight. Both of these works used ray-tracing through the Millennium Simulation \citep[][hereafter MS]{springel05,hilbert09} in order to obtain $P(\kext)$. For lines of sight which are either underdense or of common density, G13 found that the residual uncertainty $\sigma_{\kext}$ on the external convergence can be reduced to $\lesssim 0.03$, which corresponds to an uncertainty on time delay distance and hence $H_0$ comparable to that arising from the mass model of the deflector and its immediate environment. Furthermore, \citet{collett13} considered a reconstruction of the mass distribution along the line of sight using a galaxy halo model. They convert the observed environment around a lens directly into an external convergence, after calibrating for the effect of dark structures and voids by using the MS.

We have collected sufficient observational data to implement these techniques for the case of \hequad . We choose to adopt the G13 approach, with several improvements. We first aim to understand and account for various sources of error in our observational data for \hequad , as well as that of CFHTLenS \citep{heymans12}, which we choose as our control field. Second, we incorporate our understanding of these uncertainties into the simulated catalogues of the MS, in order to ensure a realistic  estimate of $P(\kext)$. Third, we use the MS to test the robustness of this estimate for simulated fields of similar under/overdensity.

This paper is organized as follows. In Section \ref{section:data} we present the relevant observational data for \hequad\, and its reduction. In Section \ref{section:CFHT} we present an overview of our control field, CFHTLenS. In Section \ref{section:lens} we present our source detection, classification, photometric redshift and stellar mass estimation, carefully designed to match the CFHTLenS fields. In Section \ref{section:overdensity} we present our technique to measure weighted galaxy count ratios for \hequad , by accounting for relevant errors. In Section \ref{section:kappa} we use ray-tracing through the MS in order to obtain $P(\kext)$ for the measured ratios, and present our tests for robustness. We present and discuss our results in Section \ref{section:discuss}, and we conclude in Section \ref{section:concl}. We present additional details in the Appendix.

The current work represents Paper III (hereafter H0LiCOW Paper III) in a series of five papers from the H0LiCOW collaboration, which together aim to obtain an accurate and precise estimate of $H_0$ from a comprehensive modelling of \hequad . An overview of this collaboration can be found in H0LiCOW Paper I (Suyu et al., submitted), and the derivation of $H_0$ is presented in H0LiCOW Paper V (Bonvin et al., submitted).  

Throughout this paper, we assume the MS cosmology, $\Om=0.25$, $\OL=0.75$, $h=0.73$, $\sigma_8 = 0.9$.\footnotemark \footnotetext{We estimate the impact of using a different cosmology in Section \ref{section:cosm_depend}.}
We present all magnitudes in the AB system, where we use the following conversion factor between the Vega and the AB systems: $J_\mathrm{AB} = J_\mathrm{Vega} + 0.91$, $H_\mathrm{AB} = H_\mathrm{Vega} + 1.35$ and $K_{s\ \mathrm{AB}} = K_{s\ \mathrm{Vega}} + 1.83$\footnotemark \footnotetext{Results based on the MOIRCS filters, available at \url{http://www.astro.yale.edu/eazy/filters/v8/FILTER.RES.v8.R300.info.txt}}. We define all standard deviations as the semi-difference between the 84 and 16 percentiles.


\section{Data reduction and calibration}\label{section:data}

In order to characterize the \hequad \ field, we require a catalogue of galaxy properties, such as galaxy redshifts and stellar masses. To this end, we have obtained multiband, wide-field imaging observations of \hequad , from ultraviolet to near/mid-infrared wavelengths. The observations are detailed in Table \ref{tab:data}, and were obtained with the Canada-France-Hawaii Telescope (CFHT; PI. S. Suyu), the Subaru Telescope (PI. C. Fassnacht), and the Gemini North Telescope (PI. C. Fassnacht). We also use archival Spitzer Telescope data (PI. C. Kochanek, Program ID 20451). In addition, we make use of a number of secure spectroscopic redshifts (374 and 43 objects inside a $\sim 17\arcmin$ and $2\arcmin$-radius circular aperture, respectively, not counting the lens itself), obtained with the Magellan 6.5m telescope \citet{momcheva06, momcheva15}, the VLT (PI: Sluse), the Keck Telescope (PI: Fassnacht), and the Gemini Telescope (PI: Treu; see H0LiCOW Paper II for details on the spectroscopic observations). Those data provide a spectroscopic identification of $\sim$ 90\% ($\sim$ 60\%) of the galaxies down to $i=21$\,mag ($i=22\,$mag)  within a radius of $2\arcmin$ of the lens, namely the maximum radius within which we calculate weighted number counts in this work (see Fig. 3 of H0LiCOW Paper II for spectroscopic completeness as a function of radius/magnitude). 

\begin{table*}
 \centering
 \begin{minipage}{155mm}
  \caption{Summary of observations}
  \begin{tabular}{@{}lllllll@{}}
  \hline 
Telescope/Instrument & FOV [$\arcmin$]/scale [$\arcsec$] & Filter & Exposure [sec] & Airmass & Seeing [$\arcsec$] & Observation date \\ 
 \hline
CFHT/MegaCam & $58 \times 56$/0.187 & $u$ & $41 \times 440$ & 1. & $\sim 0.8$ & 2014 Aug. 31 - Sep. 2 \\
Subaru/Suprime-Cam & $34 \times 27$/0.200 & $g$ & $5 \times 120$ & $\sim1.7$ & $\sim 0.7$ & 2014 Mar. 1 \\
Subaru/Suprime-Cam & $34 \times 27$/0.200 & $r$ & $16 \times 300$ & $\sim1.4$ & $\sim 0.7$ & 2014 Mar. 1 \\
Subaru/Suprime-Cam & $34 \times 27$/0.200 & $i$ & $5 \times 120$ & $\sim2.0$ & $\sim 0.8$ & 2014 Mar. 1 \\
Gemini North/NIRI & $3.4 \times 3.4$/0.116 & $J$ & $44 \times 42.2$ & $1.2 - 1.3$ & $\sim 0.4$ &  2012 Aug. 22 \\
Subaru/MOIRCS & $4 \times 7$/0.116 & $H$ & $12 \times 78$ & $1.7 - 2.1$ & $\sim 0.7$ &  2015 Apr. 1 \\
Gemini North/NIRI & $3.4 \times 3.4$/0.116 & $K_s$ & $32 \times 32.2$ & $1.2 - 1.3$ & $\sim 0.4$ &  2012 Aug. 22 \\
Spitzer/IRAC & $5.2 \times 5.2$/0.6 & 3.6 & $72 \times 30$ & - & - & 2006 Feb. 8, 2006 Sep. 20 \\
Spitzer/IRAC & $5.2 \times 5.2$/0.6 & 4.5 & $72 \times 30$ & - & - & 2006 Feb. 8, 2006 Sep. 20 \\
Spitzer/IRAC & $5.2 \times 5.2$/0.6 & 5.8 & $72 \times 30$ & - & - & 2006 Feb. 8, 2006 Sep. 20 \\
Spitzer/IRAC & $5.2 \times 5.2$/0.6 & 8.0 & $72 \times 30$ & - & - & 2006 Feb. 8, 2006 Sep. 20 \\
\hline
\end{tabular}
\\ 
{\footnotesize For NIRI, where the instrument field of view is just $2\arcmin \times 2\arcmin$, ``FOV'' refers to the effective field of view on the sky, after dithering. For IRAC, the filters denote the effective wavelengths in $\mu$m.}
\label{tab:data}
\end{minipage}
\end{table*}

We reduced the imaging data using standard reduction techniques. We obtained the CFHT MegaCam \citep{boulade03} and Spitzer IRAC \citep{fazio04} data already pre-reduced and photometrically calibrated. We used Scamp \citep{bertin06} to achieve consistent astrometric and photometric calibration, and Swarp \citep{bertin02} to resample the data on a $0.2\arcsec$ pixel scale, using a tangential projection. This is the native pixel scale of Subaru Suprime-Cam \citep{kobayashi00}, and the largest among the available data, with the exception of IRAC ($0.600\arcsec$ pixel scale). 

We reduced the Subaru MOIRCS \citep{suzuki08,ichikawa06} data using a pipeline provided by Ichi Tanaka, based on IRAF\footnotemark  \footnotetext{IRAF is distributed by the National Optical Astronomy Observatory, which is operated by the Association of Universities for Research in Astronomy (AURA) under cooperative agreement with the National Science Foundation.}.
For the Gemini NIRI \citep{hodapp03} and Subaru MOIRCS data we calibrated the photometry using 2MASS stars in the field of view (FOV). For Subaru Suprime-Cam, we used observations of an SDSS star field, taken the same night. We excluded stars with nearby companions that can affect the SDSS photometry, and used color transformations provided by Yagi Masafumi (private communication; also described in \citet{yagi13a,yagi13b}), in order to calibrate the photometry to the AB system. We corrected for galactic and atmospheric extinction following  \citet{schlafly11} and \citet{buton12}, respectively. We present our strategy for source detection, classification, redshift and stellar mass estimation, in Section \ref{section:lens}.

\section{The control field: CFHTLenS}\label{section:CFHT}

In order to apply the weighted number counts technique, we need a control field against which to determine an under/overdensity. We require the field to be of a suitable depth, as well as larger in spatial extent than the $\sim0.06 \deg^2$ field used by \citet{fassnacht11}, or the $1.21 \deg^2$ Cosmic Evolution Survey \citep[COSMOS]{scoville07a}, which is known to be overdense \citep[e.g.,][and references within]{fassnacht11}. The field should consist of several fields spread across the sky, in order to account for sample variance, and should also contain high to medium resolution, well-calibrated multiband data for object classification, and to infer photometric redshifts and stellar masses reliably.

Such a field is provided by the wide component of the CFHT Legacy Survey \citep[CFHTLS;][]{gwin12}. It consists of $ugriz$ imaging over four distinct contiguous fields: W1 ($\sim63.8$ deg$^2$ ), W2 ($\sim 22.6$ deg$^2$ ), W3 ($\sim 44.2$ deg$^2$ ) and W4 ($\sim 23.3$ deg$^2$), with typical seeing $\sim 0.7\arcsec$ in $i$-band. The data have been further processed, and are available in catalogue form from CFHTLenS \citep{heymans12}. We provide here a summary of the CFHTLenS data quality and products that are relevant to our analysis. CFHTLenS reaches down to $24.54 \pm 0.19$ $5\sigma$ limiting magnitude in a $2.0\arcsec$ aperture in the deepest band, $i$ \citep{erben13}. The photometry has been homogenized through matched and gaussianised point-spread functions (PSFs) \citep{hildebrandt12}, leading to well-characterized photometric redshifts. The CFHTLenS catalogue includes best-fit photometric redshifts derived with BPZ \citep{benitez00}, and best-fit stellar masses computed with Le PHARE \citep{ilbert06}.
The final product has a spectroscopic to photometric redshift scatter $\sigma_{|z_\mathrm{spec} - z_\mathrm{phot}|/(1+z_\mathrm{spec})}$ of $\lesssim 0.04$ for $i < 23$ ($\lesssim0.06$ for $i < 24$). The outlier fraction\footnotemark 
\footnotetext{The outliers are defined as galaxies with $|z_\mathrm{spec} - z_\mathrm{phot}|/(1+z_\mathrm{spec})<15$} is $\lesssim 5\%$ for $i < 23$ ($\lesssim15\%$ for $i < 24$) \citep{hildebrandt12}. 

The object detection and measurement are summarized by \citet{erben13}: SExtractor \citep{bertin96} is run six times in dual-image mode. In five of the runs, the detection image is the deeper image band ($i$), and the measurement images are the PSF-matched images in each of the five bands; in the sixth run, the measurement image is the original lensing band image. This last run is performed to obtain total magnitudes (SExtractor quantity MAG\_AUTO) in the deepest band, whereas the first five runs yield accurate colours based on isophotal magnitudes (MAG\_ISO).

The galaxy-star classification is summarized by \citet{hildebrandt12}, who also estimate its uncertainty, quantified in terms of incompleteness and contamination, based on a comparison with spectroscopic data from the VVDS F02 \citep[][reaching down to $i=24$ mag]{lefevre05} and VVDS F22 \citep{garilli08} surveys. In brief, for
$i < 21$, objects with size smaller than the PSF are classified as
stars. For $i > 23$, all objects are classified as galaxies. In the range
$21< i <23$, an object is defined as a star if its size is smaller than the PSF, and in addition $\chi^2_\mathrm{star} < 2.0 \chi^2_\mathrm{gal}$, where $\chi^2$ is the best-fitting goodness-of-fit $\chi^2$ from the galaxy and star libraries given by Le PHARE.

\section{Measuring physical properties of galaxies}\label{section:lens}

\subsection{Detecting and measuring sources with SExtractor}\label{section:sextractor}

\begin{figure*}
\includegraphics[width=165mm]{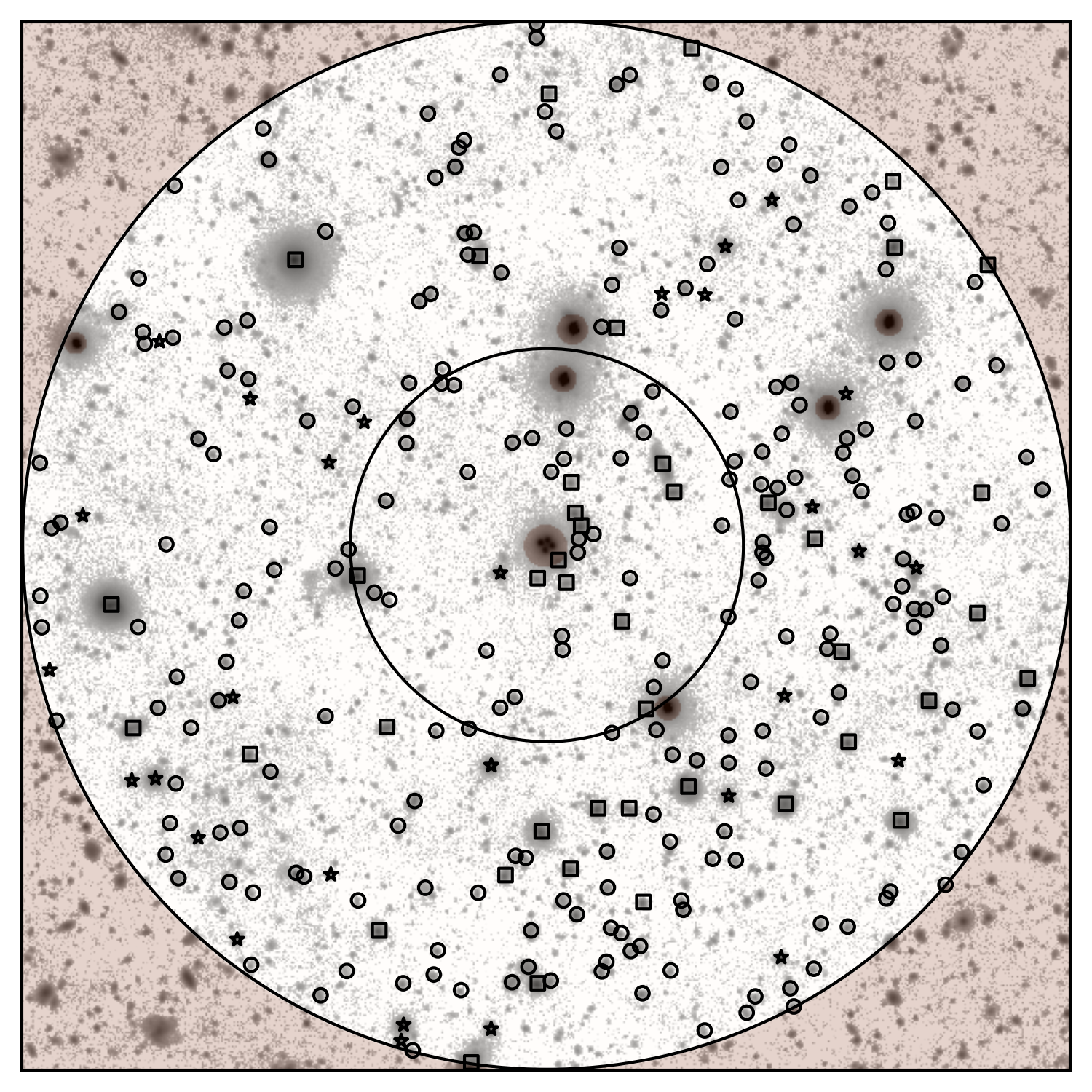}
\caption{$4\arcmin \times 4\arcmin$ FOV for \hequad\ in the deepest band, $r$. North is up and East is to the left. The $i < 24$ objects identified by SExtractor inside a $120\arcsec$ aperture are marked: star symbols for stars, circles for galaxies without spectroscopic redshift, and squares for galaxies with spectroscopic redshift. \hequad\ is at the center of the field. Brown regions represent masks outside the aperture, around the lensed system, and around bright, saturated stars. The two concentric black circles mark the $45\arcsec$ and $120\arcsec$ apertures, respectively. The nearest galaxy to the center, towards SE, is located inside the mask, as it is modelled explicitly in H0LiCOW Paper IV. For an extended FOV in $i$-band, see H0LiCOW Paper II.} 
\label{fig:sex}
\end{figure*}

In order to avoid introducing biases in measuring weighted number counts, it is important to adopt detection, measuring and classification techniques for the \hequad\ field that are as close as possible to those of CFHTLenS, while also assessing the similarities between the two datasets.

The \hequad\ $ugri$ data are similar in terms of seeing to those from CFHTLenS (Table \ref{tab:data}). The pixel scales of the two datasets differ by only $6.5\%$. In terms of depth, the limiting magnitude of the \hequad\ data in $i$-band, following the definition in \citet{erben13}\footnotemark  , is $24.55\pm0.17$, thus virtually indistinguishable from the counterpart band in CFHTLenS (Section \ref{section:CFHT}). The limiting magnitudes in the other bands are, respectively, $25.55\pm0.06$ ($u$), $25.43\pm0.20$ ($g$), $25.94\pm0.28$ ($r$), $22.71\pm0.13$ ($J$), $21.20\pm0.28$ ($H$), $21.82\pm0.28$ ($K_s$), and can be compared with the available counterparts in Table 1 of \citet{erben13}. In particular, our deepest image ($r$-band) is $\sim1$ mag deeper than the CFHTLenS $r$-band.
\footnotetext{$m_\mathrm{lim} = \mathrm{ZP} - 2.5 \log\left(5 \sqrt{N_\mathrm{pix}}\sigma_\mathrm{sky}\right)$, where ZP is the magnitude zero-point, $N_\mathrm{pix}$ is the number of pixels in a circle with radius 2.0\arcsec, and $\sigma_\mathrm{sky}$ is the sky-background noise variation. We derive the uncertainty as the standard deviation of the values in 10 empty regions across the frame.} 

To infer accurate photometry, we matched the PSFs in the $griJHK_s$ images to that in the $u$ band, which has the largest seeing. We combined bright, unsaturated stars across the field of view in each band, in order to build their PSFs. We replaced the noisy wings with analytical profiles, and computed convolution kernels using the Richardson-Lucy deconvolution algorithm \citep{richardson72,lucy74}.

Our primary region of interest is a $4\arcmin \times 4\arcmin$ area around \hequad\ , as for this area we have (for the most part) uniform coverage in all bands, including IRAC. However, it is important to also consider a larger area, in order to use as many spectroscopically observed galaxies as possible for calibrating photometric redshifts. In addition, a wider area is necessary for identifying groups/clusters (H0LiCOW Paper II), and performing a weak lensing analysis (Tihhonova et al., in prep.). As a result, we are also interested in the whole coverage of the $ugri$ frames. 

Before using SExtractor in a similar way to CFHTLenS on the $4\arcmin \times 4\arcmin$ images, we masked bright stars that are heavily saturated in $r$-band. We found that by fitting and subtracting a Moffat profile to these bright stars, we can reduce the contamination of nearby objects by the bright stars, and improve the detection parameters; this minimizes the area that needs to be masked in the $r$-band, but which is unaffected in most of the other bands. We convolve the masks with a narrow gaussian, in order smooth their edges, which would otherwise produce spurious detections. We also set a mask of $5\arcsec$ radius around the \hequad\ system itself, in order to account for the fact that the external convergence of the most nearby galaxy is accounted for explicitly in the lens mass modeling in H0LiCOW Paper IV (Wong et al., submitted). 

\begin{figure}
\includegraphics[width=90mm]{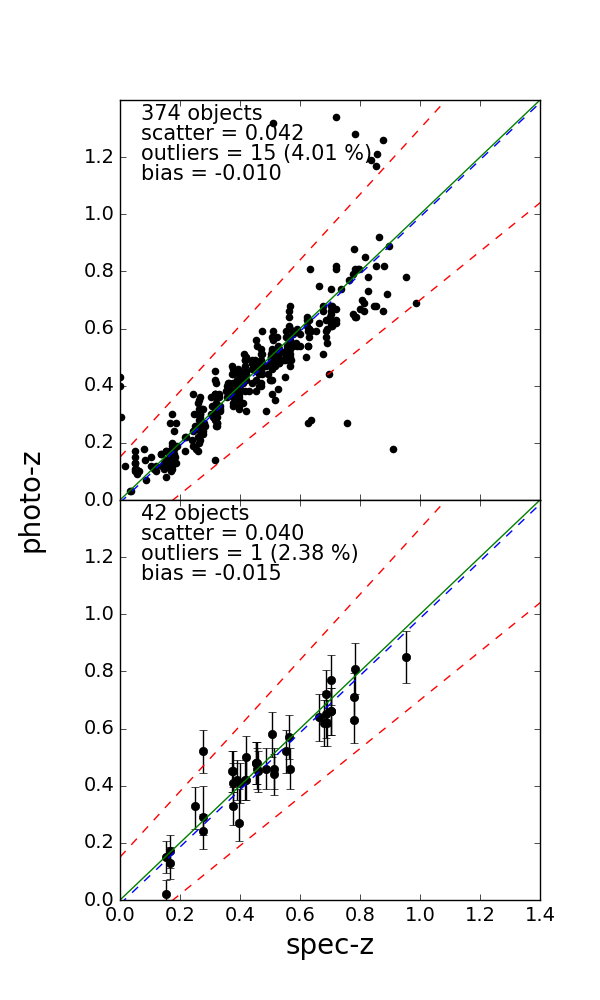}
\caption{Comparison of spectroscopic and photometric (BPZ) redshifts for all galaxies with robust spectroscopic redshifts within the Suprime-Cam FOV \textit{(left, ugri)}, as well as for the galaxies within $120\arcsec$ \textit{(right, ugriJHK)}. The blue dashed line represents the best-fit offset. We define the outliers, located outside the red dashed line, as $|z_\mathrm{spec} - z_\mathrm{phot}|/(1+z_\mathrm{spec})>0.15$, and mark this with red dashed lines. On the bottom plot, error bars refer to $1\sigma$ uncertainties determined with BPZ.
\label{fig:specz}}
\end{figure}

Despite our $r$-band being deeper, given the fact that CFHTLenS performed detections in $i$, and the similarity of our $i-$band frame to the CFHTLenS $i$-band, we first performed detections in the unconvolved (pre-PSF matching) $i$ image. For this, we ran SExtractor with the same detection parameters used by CFHTLenS (Jean Coupon, private communication). The purpose of this run is to estimate total magnitudes MAG\_AUTO in this band, which we use when performing magnitude cuts at our faint threshold. However, for the purpose of extracting reliable photometry to be used for photometric redshift and stellar mass estimation, since measurements are expected to be more reliable in $r$-band (with an exception being around bright objects, which appear brighter than in $i$), we also perform detections in this band, using optimized SExtractor detection parameters. As for measurements, we perform them as described for CFHTLenS in Section \ref{section:CFHT}. We infer final MAG\_ISO magnitudes, corrected for total magnitude, following CFHTLenS, as MAG\_ISO$_x$ + (MAG\_AUTO$_r$  $-$ MAG\_ISO$_r$), where the subscript refers to the measurement band ($x = u,g,r,i,J,H,K_s$). We make an exception for $\sim 17\%$ of objects, which have a SExtractor flag indicative of unreliable MAG\_AUTO, and for which we use replace MAG\_AUTO with MAG\_ISO instead. For the FOV outside $4\arcmin \times 4\arcmin$, which is used for separate purposes by H0LiCOW Paper II and Tihhonova et al., in prep. we performed all detections in the $r$-band only. We find that galaxies with $i\lesssim24$ mag are typically detected in all bands, with the exception of $18\%$ in $JK_s$, where the spatial coverage is also reduced, and $\sim6\%$ in $u$-band.

We use T-PHOT \citep{merlin15} to extract MAG\_ISO magnitudes, and thus measure colors between optical and IRAC filters, as the latter have vastly different pixel scale and PSFs. For this, we use the $r$-band image as position and morphology prior. Finally, we apply the same star-galaxy classification used by CFHTLenS.

Table \ref{tab:phot} compiles the $i< 23$ galaxies detected in a $45\arcsec$-radius aperture around \hequad , along with their measured photometry. The $i< 24$ galaxies in a $120\arcsec$-radius aperture can be found in the accompanying online material, and are marked on the color-combined image in Figure \ref{fig:sex}.

\subsection{Galaxy-star separation, redshifts and stellar masses}\label{section:physical}

Using the PSF-matched photometry measured with SExtractor, we infer photometric redshifts and stellar masses, which we will later use as weights. We further calibrate our magnitudes by finding the zero points which minimize the scatter between photometric and spectroscopic redshifts of the $17<i<23$ mag galaxies with available spectroscopy. 
Finally, we perform a robust galaxy-star classification using morphological as well as photometric information. For measuring redshifts, we primarily use BPZ, which was also employed by CFHTLenS. However, we also use EAZY \citep{brammer08}, to assess the dependence on a particular code/set of templates.

For the purpose of estimating photometric redshifts we ignore the IRAC channels, as e.g. \citet[][]{hildebrandt10} note that the use of currently available mid-IR templates degrade rather than improve the quality of the inferred redshifts. For both BPZ and EAZY, we obtained the best results when using the default set of templates (CWW+SB and a linear combination of principal component spectra, respectively), with the default priors. Figure \ref{fig:specz} compares the available spectroscopic redshifts with the inferred photometric redshift for the $ugriJHK_s$ and $ugri$ filters, and galaxies with $i<24$ mag. There is negligible bias, and the scatter/outlier fractions are comparable to or smaller than the ones for CFHTLenS (Section \ref{section:CFHT}). In addition, Figure \ref{fig:photoz} compares the BPZ- and EAZY-estimated redshifts, for the $i<24$ galaxies inside the $4\arcmin \times 4\arcmin$ region around  \hequad , showing a good overall match.

For estimating stellar masses, we followed the approach by \citet{erben13}, which was also used to produce the CFHTLenS catalogues. This uses templates based on the stellar population synthesis package of \citet{bruzual03}, with a \citet{chabrier03} initial mass function (see \citet{velander14} for additional details), and fits stellar masses with Le PHARE, at fixed redshift. We performed the computation twice, without and with using the IRAC photometry. In the latter case, we boosted the photometric errors to account for the template error derived by \citet{brammer08}. We find only small scatter ($\sim0.05$ in $\log M_\star$) and no bias, in agreement with the results of \citet{ilbert10} for a similar redshift range. The resulting redshifts and the stellar masses are given in Table \ref{tab:zmstar}. We used the median of the mass probability distribution as our estimate, except for a few percent of galaxies where Le PHARE fails to give a physical estimate for this, and we use the best-fit value instead. This is also the case for the CFHTLenS catalogues, where we recomputed stellar masses in order to fix the $\sim 6\%$ of objects with missing estimates. In fact, we recomputed stellar masses for the whole CFHTLenS catalogues, in order to use the same cosmology employed by the MS. 

Finally, following the recipe from Section \ref{section:CFHT}, we performed a galaxy-star classification. As described in more detail by \citet{hildebrandt12}, we estimated the PSF size as the $3 \sigma$ upper cut half light radius estimated by SExtractor in $r$-band, and we used all available bands when computing the goodness-of-fit. Comparing to the available spectroscopic data, we find that all spectroscopically-confirmed galaxies are correctly classified as galaxies, whereas three spectroscopically-confirmed stars, with blended galaxy contaminants, are incorrectly classified as galaxies. We therefore removed them.

\section{Determining line-of-sight under/overdensities using weighted number counts}\label{section:overdensity}

\subsection{Description of the technique}\label{section:ratiosdescript}

\citet{fassnacht11} computed lens field overdensities as galaxy count ratios, by first measuring the mean number counts in a given aperture through their control field, and then dividing the counts in the same aperture around the lens to the mean, i.e. $\zeta_\mathrm{gal}\equiv N_\mathrm{gal}/\overline{N_\mathrm{gal}}$. The situation is more complicated for us, because 1) we are interested in using weights dependent on the particular galaxy position inside the aperture, and 2) the CFHTLenS control fields contain a large fraction of masks throughout. These masked areas are due to luminous halos around saturated stars, asteroid trails, flagged pixels etc. \citep{erben13}.

\begin{figure}
\includegraphics[width=83mm]{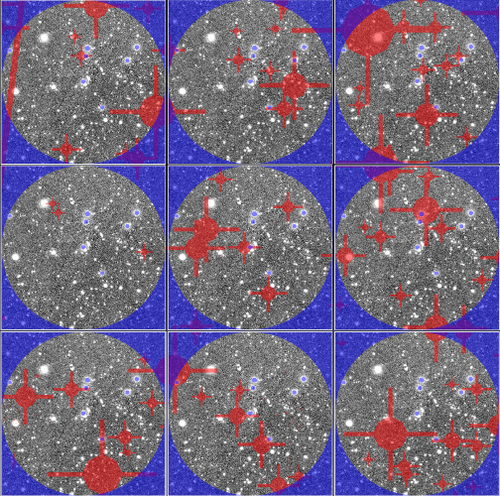}
\caption{Schematic of the way masking is applied when matching the \hequad\ and various CFHTLenS subfields, on a grid. The  $2\arcmin$-radius $i$-band frame (gray), masks around bright stars and outside the aperture of \hequad\ (blue), and masks in the CFHTLenS fields (red) are depicted. Only the gray area which is not covered by any masks is used.
\label{fig:mask}}
\end{figure}

Therefore, to implement our galaxy weighting schemes, we first divide each of each of the W1-W4 CFHTLenS fields into a two-dimensional, contiguous grid of cells, of the same size as the apertures we consider around \hequad . We apply the CFHTLenS masks, at their particular position inside the cell, to the \hequad\ field as well. Thus, when measuring weighted counts, we test whether each galaxy in the \hequad\ field is located at a position which is covered by a mask in a particular cell. Conversely, we also test whether a galaxy in the cell is covered by a mask in the \hequad \ field. This technique is depicted in Figure \ref{fig:mask}. 

We divide the weighted counts measured around \hequad\  to those measured in the same way around the center of each of the cells in the CFHTLenS grid, and consider the median of these divisions as our estimate of the overdensity. We justify the use of the median in Section \ref{section:overdensities} and Appendix~\ref{section:MSdetails}. Formally, $\zeta_\mathrm{gal}$ then becomes $\overline{\zeta_\mathrm{gal}^\mathrm{WX}} \equiv \mathrm{median}\left(  N_\mathrm{gal}^{\mathrm{lens, mask}_\mathrm{i}}/N_\mathrm{gal}^\mathrm{WX,i}\right)$, where $X=1,...,4$ and $i$ spans the number of cells in a CFHTLenS field. Following the notation in G13, we generalize from number counts to weighted counts by replacing $N_\mathrm{gal}$ with $W_q=\sum_{i=1}^{N_\mathrm{gal}}q_i$, where $q$ refers to a particular type of weight. Therefore $\zeta_\mathrm{gal}$ generalizes to $\zeta_q$. 

Following G13 we adopt these weights: $q_\mathrm{gal}=1$, i.e. simple galaxy counting; $q_{M^n_\star}=M^n_\star\  (n=1,2,3)$, i.e. summing up powers of galaxy stellar masses; and $q_z = \zs \cdot z - z^2$. In addition, we also consider weights incorporating the distance to the lens/center of the field: $q_{1/r}$, $q_{M^n_\star/r}$, and $q_{z/r}$, as well as the weighted counts $W_{M^n_{\mathrm{\star,rms}}}= \sqrt[n]{\sum_{i=1}^{N_\mathrm{gal}} M^n_{\star,i}}$ and $W_{M^n_\star/r_\mathrm{,rms}}, \ (n=2,3)$. 


In addition to the weights from G13, we define an additional weight, $M_\star/r^n$, where $n=2,3$ corresponds to the tidal and the flexion shift, respectively, of a point mass, as defined in \citet{mccully16}. We have simplified the definition of these two quantities, by removing the explicit redshift dependence. This is because the lensing convergence maps of the MS are not designed to account for this dependence \citep{hilbert09}. 
 Another weight, $\sqrt{M_\star}/r$, corresponds to the convergence produced by a singular isothermal sphere. We supplement this with a final related weight, $\sqrt{M_h}/r$, where $M_h$ stands for the halo mass of the galaxy, derived from the stellar mass by using the relation of \citet{behroozi10}. 

\onecolumn
\begin{landscape}
\scriptsize
\begin{longtable}{cccccccccccccc}
\caption{Photometric properties of the $i\leq23$ galaxies inside $45\arcsec$ of \hequad } \\ 
  \hline 
RA  & DEC & $i_\mathrm{tot}$ & $u$ & $g$ & $r$ & $i$ & $J$ & $H$ & $K_s$ & $3.6\mu m$ & $4.5\mu m$ & $5.7\mu m$ & $7.9\mu m$  \\ 
 \hline
69.57430 & $-12.28945$ & 18.64 &  $22.55 \pm 0.02$ &  $20.93 \pm 0.01$ &    $19.45 \pm 0.01$  &   $18.73 \pm 0.01$  &   $18.01 \pm 0.01$ &    $17.71 \pm 0.01$ &    $17.48 \pm 0.01$ &       $17.90 \pm 0.23$ &     $18.21 \pm 0.27$ &     $18.36 \pm 0.30 $ &     $19.19 \pm 0.35 $ \\
69.55442 & $-12.28233$ & 20.12 &  $22.47 \pm 0.02$ &  $21.44 \pm 0.01$ &    $20.60 \pm 0.01$  &   $20.16 \pm 0.01$  &   $19.32 \pm 0.01$ &    $18.77 \pm 0.01$ &    $18.53 \pm 0.01$ &       $18.93 \pm 0.23$ &     $18.80 \pm 0.27$ &     $18.96 \pm 0.30 $ &     $17.53 \pm 0.35 $ \\
69.55975 & $-12.28627$ & 20.21 &  $21.89 \pm 0.01$ &  $21.28 \pm 0.01$ &    $20.49 \pm 0.01$  &   $20.23 \pm 0.01$  &   $19.61 \pm 0.01$ &    $19.38 \pm 0.02$ &    $19.15 \pm 0.01$ &       $19.56 \pm 0.23$ &     $19.51 \pm 0.27$ &     $19.80 \pm 0.30 $ &     $18.73 \pm 0.35 $ \\
69.56122 & $-12.28845$ & 20.43 &  $22.81 \pm 0.02$ &  $22.19 \pm 0.01$ &    $21.37 \pm 0.01$  &   $20.50 \pm 0.01$  &   $19.40 \pm 0.01$ &    $18.97 \pm 0.01$ &    $18.72 \pm 0.01$ &       $18.65 \pm 0.23$ &     $19.08 \pm 0.27$ &     $19.27 \pm 0.30 $ &     $18.96 \pm 0.35 $ \\
69.55652 & $-12.27911$ & 20.73 &  $23.41 \pm 0.04$ &  $22.37 \pm 0.01$ &    $21.28 \pm 0.01$  &   $20.81 \pm 0.01$  &   $19.99 \pm 0.01$ &    $19.58 \pm 0.02$ &    $19.42 \pm 0.01$ &       $19.86 \pm 0.23$ &     $19.99 \pm 0.27$ &     $20.29 \pm 0.30 $ &     $19.81 \pm 0.35 $ \\
69.56013 & $-12.28546$ & 21.03 &  $23.77 \pm 0.04$ &  $22.76 \pm 0.01$ &    $21.59 \pm 0.01$  &   $21.05 \pm 0.01$  &   $20.18 \pm 0.01$ &    $19.71 \pm 0.02$ &    $19.60 \pm 0.01$ &       $19.88 \pm 0.23$ &     $20.09 \pm 0.27$ &     $20.32 \pm 0.30 $ &     $19.48 \pm 0.35 $ \\
69.55710 & $-12.29236$ & 21.04 &  $22.66 \pm 0.02$ &  $22.20 \pm 0.01$ &    $21.60 \pm 0.01$  &   $21.11 \pm 0.01$  &   $20.75 \pm 0.01$ &    $20.36 \pm 0.04$ &    $20.22 \pm 0.01$ &       $20.37 \pm 0.23$ &     $20.74 \pm 0.27$ &     $20.68 \pm 0.31 $ &     $20.69 \pm 0.37 $ \\
69.55370 & $-12.28414$ & 21.19 &  $23.61 \pm 0.04$ &  $22.81 \pm 0.01$ &    $21.59 \pm 0.01$  &   $21.18 \pm 0.01$  &   $20.37 \pm 0.01$ &    $19.96 \pm 0.02$ &    $19.86 \pm 0.01$ &       $20.28 \pm 0.23$ &     $20.35 \pm 0.27$ &     $20.50 \pm 0.30 $ &     $19.78 \pm 0.35 $ \\
69.57109 & $-12.27948$ & 21.24 &  $23.34 \pm 0.03$ &  $22.50 \pm 0.01$ &    $21.61 \pm 0.01$  &   $21.29 \pm 0.01$  &   $20.74 \pm 0.01$ &    $20.44 \pm 0.03$ &    $20.20 \pm 0.01$ &       $20.79 \pm 0.23$ &     $20.84 \pm 0.27$ &     $20.62 \pm 0.30 $ &     $20.45 \pm 0.36 $ \\
69.55553 & $-12.29793$ & 21.67 &  $23.27 \pm 0.03$ &  $22.83 \pm 0.01$ &    $22.06 \pm 0.01$  &   $21.81 \pm 0.01$  &   $21.54 \pm 0.02$ &    $22.99 \pm 0.40$ &    $21.58 \pm 0.03$ &       $21.71 \pm 0.23$ &     $22.14 \pm 0.28$ &     $22.41 \pm 0.44 $ &     -                 \\
69.56258 & $-12.28962$ & 21.73 &  $26.51 \pm 0.41$ &  $24.66 \pm 0.04$ &    $22.92 \pm 0.01$  &   $21.68 \pm 0.01$  &   $20.52 \pm 0.01$ &    $20.11 \pm 0.02$ &    $19.77 \pm 0.01$ &       $19.61 \pm 0.23$ &     $20.03 \pm 0.27$ &     $20.44 \pm 0.30 $ &     $20.38 \pm 0.36 $ \\
69.56038 & $-12.28351$ & 22.00 &  $27.31 \pm 0.75$ &  $24.70 \pm 0.04$ &    $23.10 \pm 0.01$  &   $22.02 \pm 0.01$  &   $20.84 \pm 0.01$ &    $20.34 \pm 0.03$ &    $20.17 \pm 0.01$ &       $19.99 \pm 0.23$ &     $20.44 \pm 0.27$ &     $20.38 \pm 0.30 $ &     $20.66 \pm 0.37 $ \\
69.56071 & $-12.28990$ & 22.16 &  $27.20 \pm 0.81$ &  $24.68 \pm 0.05$ &    $23.01 \pm 0.01$  &   $22.07 \pm 0.01$  &   $20.77 \pm 0.01$ &    $20.45 \pm 0.04$ &    $20.05 \pm 0.01$ &       $19.84 \pm 0.23$ &     $20.27 \pm 0.27$ &     $20.35 \pm 0.30 $ &     $20.76 \pm 0.38 $ \\
69.57321 & $-12.29053$ & 22.31 &  $24.04 \pm 0.06$ &  $23.68 \pm 0.02$ &    $22.69 \pm 0.01$  &   $22.23 \pm 0.02$  &   $22.12 \pm 0.03$ &    $21.45 \pm 0.09$ &    $21.61 \pm 0.03$ &       $21.88 \pm 0.23$ &     $22.22 \pm 0.28$ &     -                 &     -                 \\
69.55510 & $-12.27773$ & 22.73 &  $26.06 \pm 0.20$ &  $24.70 \pm 0.03$ &    $23.48 \pm 0.01$  &   $22.90 \pm 0.02$  &   $22.08 \pm 0.02$ &    $22.08 \pm 0.11$ &    $21.56 \pm 0.02$ &       $21.75 \pm 0.23$ &     $22.00 \pm 0.27$ &     $21.94 \pm 0.35 $ &     -                 \\
69.55990 & $-12.28712$ & 22.81 &  $24.26 \pm 0.04$ &  $23.81 \pm 0.02$ &    $23.42 \pm 0.01$  &   $22.91 \pm 0.02$  &   $22.42 \pm 0.02$ &    $22.26 \pm 0.14$ &    $22.32 \pm 0.03$ &       $21.70 \pm 0.23$ &     $21.92 \pm 0.27$ &     $22.81 \pm 0.53 $ &     $20.33 \pm 0.36 $ \\
69.55017 & $-12.29207$ & 22.89 &  $26.61 \pm 0.35$ &  $26.10 \pm 0.12$ &    $24.12 \pm 0.01$  &   $22.99 \pm 0.02$  &   $21.74 \pm 0.01$ &    $20.81 \pm 0.03$ &    $20.84 \pm 0.01$ &       $20.56 \pm 0.23$ &     $21.05 \pm 0.27$ &     $20.94 \pm 0.31 $ &     -                 \\
69.56423 & $-12.28099$ & 22.93 &  $24.46 \pm 0.05$ &  $23.82 \pm 0.03$ &    $23.29 \pm 0.01$  &   $23.07 \pm 0.02$  &   $22.55 \pm 0.03$ &    $22.40 \pm 0.15$ &    $22.41 \pm 0.04$ &       $23.12 \pm 0.25$ &     $23.58 \pm 0.36$ &     $22.08 \pm 0.41 $ &     $22.05 \pm 0.63 $ \\
69.55569 & $-12.28037$ & 22.96 &  $24.27 \pm 0.04$ &  $23.81 \pm 0.02$ &    $23.74 \pm 0.01$  &   $23.34 \pm 0.03$  &   $22.65 \pm 0.03$ &    $21.70 \pm 0.09$ &    $22.34 \pm 0.04$ &       $21.21 \pm 0.23$ &     $21.06 \pm 0.27$ &     $21.15 \pm 0.32 $ &     $21.35 \pm 0.42 $ \\
69.56407 & $-12.29717$ & 22.98 &  $23.69 \pm 0.03$ &  $23.53 \pm 0.01$ &    $23.34 \pm 0.01$  &   $22.98 \pm 0.02$  &   $22.40 \pm 0.03$ &    $21.94 \pm 0.10$ &    $21.91 \pm 0.02$ &       $21.58 \pm 0.23$ &     $21.90 \pm 0.27$ &     $22.43 \pm 0.42 $ &     -                 \\
\hline
\label{tab:phot}
\end{longtable}
\vspace*{-7.0mm}
{\scriptsize The complete catalogue of $i\leq24$ galaxies inside $120\arcsec$ is available as online material, and the $ugri$ photometry for the complete Subaru/Suprime-Cam FOV is available upon request. Galaxies covered by the masks in Figure \ref{fig:sex}, except for the nearest companion inside $5\arcsec$ of \hequad , are not reported. Here $i_\mathrm{tot}$ is the SExtractor $\mathrm{MAG\_AUTO}$ with detections in $i$-band, and the rest are $\mathrm{MAG\_ISO}$ magnitudes with detections in the $r$-band, corrected by adding $\mathrm{MAG\_AUTO_r}$ $-$ $\mathrm{MAG\_ISO_r}$. Reported magnitudes are corrected for atmospheric (when necessary) and galactic extinction, but not for the zero point offsets estimated by BPZ (with the exception of $i_\mathrm{tot}$; see text). These offsets are: $\Delta u =-0.07$, $\Delta g =0.12$, $\Delta r =0.05$, $\Delta i =-0.02$, $\Delta J =-0.01$, $\Delta H =0.06$ and $\Delta K_s =0.09$. For the IRAC channels, errors include those from the EAZY template error function.}

\begin{longtable}{cccccccc|ccccccccc}
\caption{Inferred redshifts, stellar and halo masses of the $i\leq23$ galaxies inside $45\arcsec$ of \hequad} \\ 
  \hline 
RA  & DEC & sep & zspec/bpz & $z_\mathrm{16\%}$ & $z_\mathrm{84\%}$ & $\log M_\star$ & $\log M_\mathrm{halo}$ & & RA  & DEC & sep & zspec/bpz & $z_\mathrm{inf}$ & $z_\mathrm{sup}$ & $\log M_\star$ & $\log M_\mathrm{halo}$ \\
 \hline
69.57430 & $-12.28945$ & 43.88 &   0.515 &  -  	&  -   & 11.1500  & 14.0037  & & 69.56258 & $-12.28962$ &  7.87 &   0.781 &  -  	&  -   & 10.5502  & 12.8158 \\
69.55442 & $-12.28233$ & 32.48 &   0.277 &  -  	&  -   & 9.9580   & 12.1143  & & 69.56038 & $-12.28351$ & 15.47 &   0.702 &  -  	&  -   & 10.4472  & 12.6788 \\
69.55975 & $-12.28627$ &  9.05 &   0.419 &  -  	&  -   & 9.9445   & 12.1451  & & 69.56071 & $-12.28990$ &  9.71 &   0.779 &  -      & -    & 10.7420  & 13.0907 \\
69.56122 & $-12.28845$ &  4.32 &   0.782 &  -  	&  -   & 10.9000  & 13.3647  & & 69.57321 & $-12.29053$ & 40.96 &   0.48  &  0.41   & 0.55 & 9.2501   & 11.7876 \\
69.55652 & $-12.27911$ & 35.83 &   0.41  &  0.34& 0.48 & 10.1380  & 12.2910  & & 69.55510 & $-12.27773$ & 42.73 &   0.36  &  0.29   & 0.43 & 9.3728   & 11.8225 \\
69.56013 & $-12.28546$ &  9.85 &   0.457 &  -  	&  -   & 10.3990  & 12.5692  & & 69.55990 & $-12.28712$ &  7.47 &   0.64  &  0.56   & 0.72 & 9.3000   & 11.8362 \\
69.55710 & $-12.29236$ & 24.50 &   0.678 &  -  	&  -   & 9.8689   & 12.1570  & & 69.55017 & $-12.29207$ & 44.68 &   0.81  &  0.72   & 0.90 & 10.3029  & 12.5396 \\
69.55370 & $-12.28414$ & 31.57 &   0.419 &  -  	&  -   & 10.2032  & 12.3521  & & 69.56423 & $-12.28099$ & 24.77 &   0.25  &  0.19   & 0.31 & 8.5401   & 11.4583 \\
69.57109 & $-12.27948$ & 43.15 &   0.37  &  0.30& 0.44 & 9.5001   & 11.8830  & & 69.55569 & $-12.28037$ & 33.95 &   0.87  &  0.78   & 0.96 & 9.1534   & 11.8060 \\
69.55553 & $-12.29793$ & 43.84 &   0.488 &  -  	&  -   & 9.3000   & 11.8110  & & 69.56407 & $-12.29717$ & 35.52 &   1.01  &  0.91   & 1.11 & 9.6334   & 12.1157 \\
\hline
\label{tab:zmstar}
\end{longtable}
\vspace*{-7.0mm}
{\scriptsize The complete catalogue of $i\leq24$ galaxies inside $120\arcsec$ is available as online material, and that of the complete Subaru/Suprime-Cam FOV, based on $ugri$ photometry, is available upon request. Where $z_\mathrm{16\%}$ and $z_\mathrm{84\%}$ values are not given, spectroscopic redshifts are available. Photometric redshift values correspond to the peak of the probability distributions, and logarithmic mass values correspond to the medians of the probability distributions estimated with Le PHARE. The typical uncertainty given by Le PHARE (including IRAC photometry) for $M_\star$ is $\sim0.05$ dex.}
\end{landscape}
\twocolumn

 Finally, in addition to the summed weighted counts used by G13, we introduce an alternative type of weighted counts, which as we will later show, produces improved results. We refer to $W_q$ defined above as $W_q^\mathrm{sum}$, and we define $W_q^\mathrm{meds} = N_\mathrm{gal}\cdot \mathrm{median}(q_i),\ i=1, ..., N_\mathrm{gal}$. All of the weights and weighted counts defined above are summarized in Table \ref{tab:weights}. Separately from these, we will also use a supplementary constraint when selecting lines of sight from the MS: the shear value at the location of \hequad , $\gext = 0.030\pm 0.004$, as measured in H0LiCOW Paper IV for the fiducial lens model. 

Following G13, we only consider galaxies of redshift $0<z<\zs$, and for $r\leq10\arcsec$ we replace $1/r$ in all weights incorporating $1/r$ with 1/10, in order to limit the contribution of the most nearby galaxies, which are accounted for explicitly in the mass model (paper IV).  For the \hequad \ field, where available, we use spectroscopic redshifts for every galaxy, and photometric redshifts for the rest. For CFHTLenS, we impose a bright magnitude cut of $i=17.48$, corresponding to the brightest galaxy in the \hequad \ field.

The final quantities that remain to be chosen are the aperture size and depth that we consider, both for the field around \hequad , and for CFHTLenS. \citet{fassnacht11} used a single aperture of $45\arcsec$ radius and galaxies down to 24 mag in F814W (Vega-based), mainly motivated by the size and depth of the HST/ACS chip used for their observations. G13 also adopted the same aperture and depth. Using their galaxy halo-model approach to reconstruct the mass distribution along the line of sight, \citet{collett13} determined using the MS that the majority of the $\kext$ comes from galaxies inside an aperture of $2\arcmin$-radius and brighter than $i=24$ mag. Although our relative counts technique may reduce the sensitivity to the choice of aperture and depth, our observation campaigns were thus designed to reach $i=24$ over a $2\arcmin$-radius aperture in light of the \citet{collett13} results.

\begin{table}
 \centering
 \begin{minipage}{155mm}
  \caption{Types of weights and weighted counts}
  \begin{tabular}{@{}lll@{}}
  \hline 
$q$ & $W_q^\mathrm{sum}$ & $W_q^\mathrm{meds}$ \\
 \hline
1 & $N_\mathrm{gal}$ & $N_\mathrm{gal}$ \\
$z$ & $\sum_{i=1}^{N_\mathrm{gal}}\left(\zs \cdot z_i - z_i^2\right)$ & $N_\mathrm{gal}\cdot \mathrm{med}\left(\zs \cdot z_i - z_i^2\right)$ \\
$M^n_\star$ & $\sum_{i=1}^{N_\mathrm{gal}}M^n_{\star,i}$ & $N_\mathrm{gal}\cdot \mathrm{med}\left(M^n_{\star,i}\right)$ \\
$1/r$ & $\sum_{i=1}^{N_\mathrm{gal}}1/r_i$ & $N_\mathrm{gal}\cdot \mathrm{med}\left(1/r_i\right)$ \\
$z/r$ & $\sum_{i=1}^{N_\mathrm{gal}}\left(\zs \cdot z_i - z_i^2\right)/r_i$ & $N_\mathrm{gal}\cdot \mathrm{med}\left(\zs \cdot z_i - z_i^2\right)/r_i$ \\
$M^n_\star/r$ & $\sum_{i=1}^{N_\mathrm{gal}}M^n_{\star,i}/r_i$ & $N_\mathrm{gal}\cdot \mathrm{med}\left(M^n_{\star,i}/r_i\right)$ \\
$M^n_\mathrm{\star,rms}$ & $ \sqrt[n]{\sum_{i=1}^{N_\mathrm{gal}} M_{\star,i}^n}$ & $\sqrt[n]{N_\mathrm{gal}\cdot \mathrm{med}\left(M^n_{\star,i}\right)}$ \\
$M^n_\star/r_\mathrm{,rms}$ & $ \sqrt[n]{\sum_{i=1}^{N_\mathrm{gal}} M_{\star,i}^n/r_i}$ & $\sqrt[n]{N_\mathrm{gal}\cdot \mathrm{med}\left(M^n_{\star,i}/r_i\right)}$ \\
$M_\star/r^n$ & $\sum_{i=1}^{N_\mathrm{gal}}M_{\star,i}/r_i^n$ & $N_\mathrm{gal}\cdot \mathrm{med}\left(M_{\star,i}/r_i^n\right)$ \\
$\sqrt{M_\star}/r$ & $\sum_{i=1}^{N_\mathrm{gal}}\sqrt{M_{\star,i}}/r_i$ & $N_\mathrm{gal}\cdot \mathrm{med}\left(\sqrt{M_{\star,i}}/r_i\right)$ \\
$\sqrt{M_h}/r$ & $\sum_{i=1}^{N_\mathrm{gal}}\sqrt{M_{h,i}}/r_i$ & $N_\mathrm{gal}\cdot \mathrm{med}\left(\sqrt{M_{h,i}}/r_i\right)$ \\
\hline
\end{tabular}
\\ 
{\footnotesize Here ``med'' refers to the median, and $n=1,2,3$ for weights not \\ 
including ``rms'' or $r$ to powers larger than 1, $n=2,3$ otherwise.}
\label{tab:weights}
\end{minipage}
\end{table}

\begin{figure*}
\includegraphics[width=175mm]{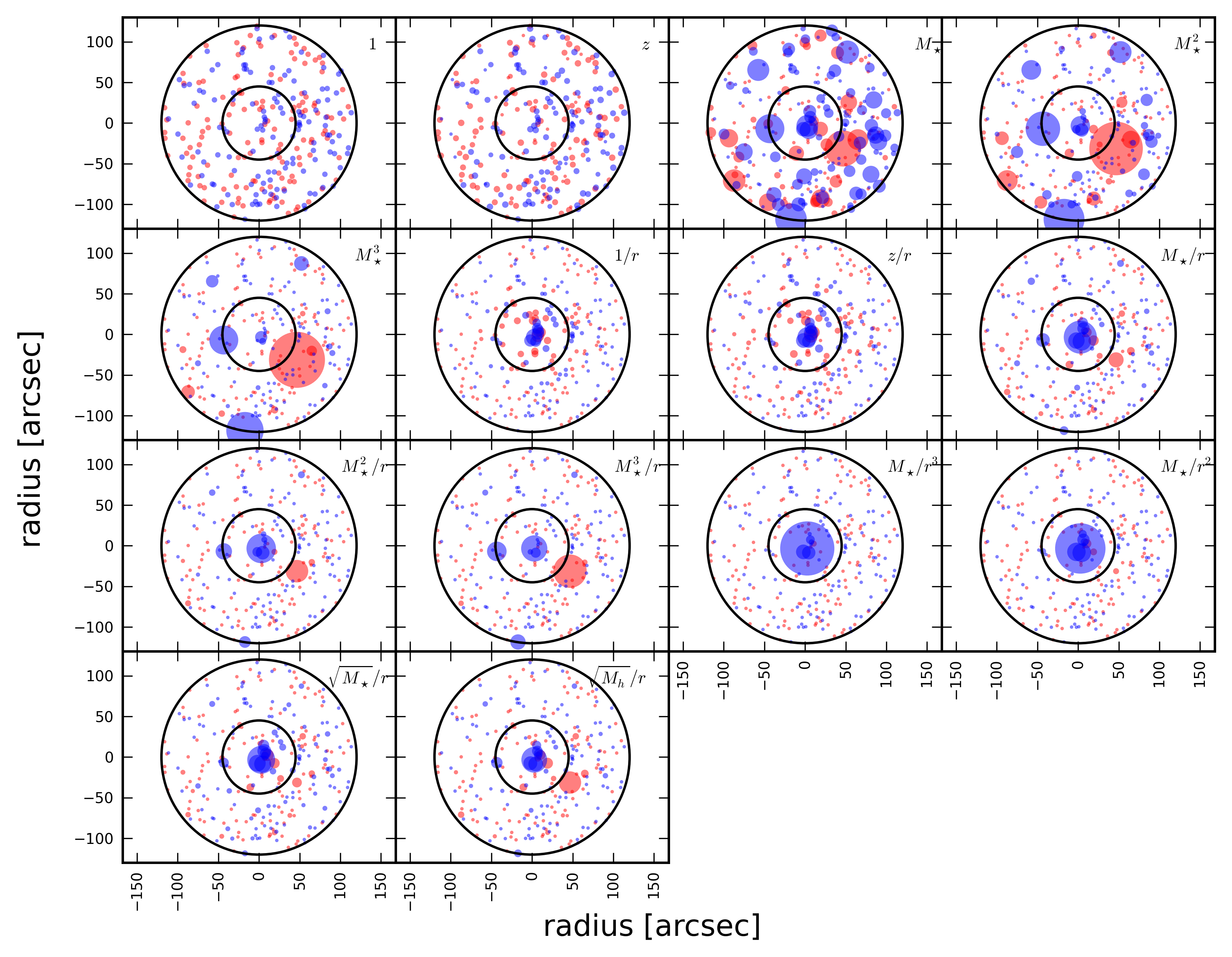}
\caption{ The relative weights of the galaxies around \hequad , represented by circles with areas proportional to their weights. Blue circles refer to  $i\leq23$ mag galaxies, whereas red circles refer to $23<i<24$ mag galaxies. A constant minimum circle radius is used for legibility. 
\label{fig:bubbles}}
\end{figure*}

Finally, in Figure \ref{fig:bubbles} we show the relative weight of each galaxy in the \hequad\ field, where we mark our magnitude and aperture limits. As designed, galaxies very close to the lens have larger weight, particularly for $q=M_\star/r^3$ and $M_\star/r^2$, as are more massive, and comparatively brighter galaxies.

\subsection{Resulting distributions for $\zeta_q$}\label{section:overdensities}

In section we present our results, regarding the distribution of overdensities.
The results are robust to different sources of systematic and random uncertainties, as we show in detail in Appendix~\ref{section:systematics}. The uncertainties discussed in the Appendix include the choice of different aperture radii ($45\arcsec$ and $120\arcsec$) and limiting magnitudes ($i<23$ and $i<24$), using CFHTLenS cells with at least 75\% or 50\% of their surface free of masks, considering the W1-W4 CFHTLenS individually in order to assess sample variance, and sampling from the inferred distribution of redshift and stellar mass for each galaxy. 

\begin{figure*}
\includegraphics[width=185mm]{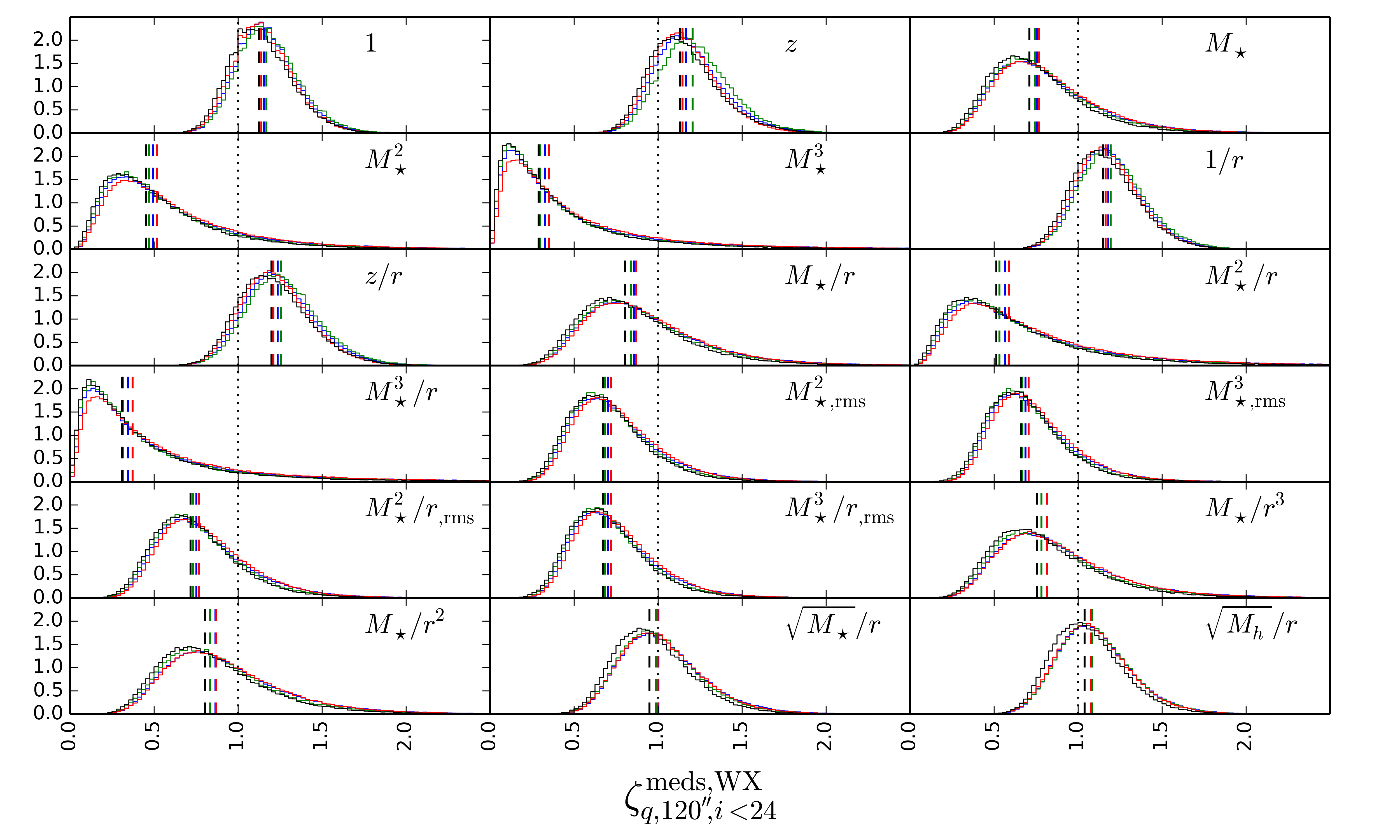}
\caption{Normalized histograms of weighted ratios for all $\zeta_q^\mathrm{meds,WX}$ weights, where $X=1\ \mathit{(blue)}, X=2\ \mathit{(green)},X=3\ \mathit{(red)},X=4\ \mathit{(black)}$, for galaxies inside a $120\arcsec$-radius aperture and $i\leq 24$. We only plot the distributions obtained from using CFHTLenS apertures with at least 75\% of their surface free of masks, as the 75\% - 50\% limit distributions appear virtually identical. The vertical dashed lines mark the medians of the distributions. 
\label{fig:12024meds}}
\end{figure*}

We plot $\zeta_q^{WX}, X=1,2,3,4$ for all weights $q$ as well as a selection of aperture radii and limiting magnitudes, in Figures \ref{fig:12024meds}, \ref{fig:4523meds} and \ref{fig:12024sum}. These are known as ratio distributions, or, more approximately, inverse gaussian distributions. There are two reasons why we take the medians of these distributions as an estimate of the field under/overdensity, $\overline{\zeta_q^{WX}}\equiv\mathrm{median}\left({\zeta_q^{WX}} \right)$, instead of the mean. First, because the median is robust to the long tails displayed by some of the distributions, whereas the use of means would imply that the field is of unphysically large overdensity. Second, so that we can use a numerical approximation which decreases significantly the computation time when estimating weighted count ratios in the MS (see Section \ref{section:MSdetails} for details). This approach is much faster and more robust than clipped averages.

By comparing the $\zeta^{WX}_q$ distributions for different magnitude and aperture limits (Figures \ref{fig:12024meds} and \ref{fig:4523meds}), it is apparent that the distributions corresponding to brigher limiting magnitude and smaller aperture are wider. This is due to larger Poisson noise when computing weighted counts, since fewer galaxies are included. In Figures \ref{fig:12024meds} and \ref{fig:12024sum} we show the distributions for $\zeta_q^\mathrm{meds}$ and $\zeta_q^\mathrm{sum}$, respectively. $\zeta_q^\mathrm{sum}$ shows more scatter between W1-W4, and as we will show in Section \ref{section:kappa}, it is also more noisy. It also shows more clearly that fields W1 and W3 are relatively more similar to each other, and different from W2 and W4, as expected from the fact that these two latter fields have a larger fraction of star contaminants (see Section \ref{section:systematics}). We find that distributions using cells with masked fractions $<50\%$ and $<25\%$ are very similar, at $\sim1\%$ level. The scatter in $\zeta_q^{\mathrm{meds},WX}$ for a given weight $q$ (hereafter we only consider W1 and W3, given the result above) is also very small, indicating that sample variance in CFHTLenS is not an issue. The distributions are virtually unchanged if we compute stellar masses with or without the IRAC bands, and very similar whether EAZY or BPZ are used to compute redshifts. We find the largest differences when using different SExtractor detection parameters (in particular for the deeper magnitude limit of $i<24$ mag), and when comparing the 10 distributions obtained from sampling from the redshift and stellar mass distributions of each galaxy (see Section \ref{section:systematics}). In Table \ref{tab:overdens}, we give the measured weighted ratios, where we include when computing the medians all the source of scatter discussed above. 

\begin{table*}
 \centering
 \begin{minipage}{155mm}
  \caption{Weighted galaxy count ratios $\overline{\zeta_q}$ for \hequad}
  \begin{tabular}{@{}lcccccccc@{}}
  \hline 
& $45\arcsec$ & $45\arcsec$ & $45\arcsec$ & $45\arcsec$ & $120\arcsec$ & $120\arcsec$ & $120\arcsec$ & $120\arcsec$ \\
Weight $q$ & $i < 24$ & $i < 24$ & $i < 23$ & $i < 23$ & $i < 24$ & $i < 24$ & $i < 23$ & $i < 23$  \\
& sum & meds & sum & meds & sum & meds & sum & meds  \\
 \hline
$1$               			& $1.27\pm0.05$ & $1.27\pm0.05$ & $1.35\pm0.04$ & $1.35\pm0.04$ & $1.15\pm0.04$ & $1.15\pm0.04$ & $1.23\pm0.03$ & $1.23\pm0.03$  \\ 
$z$                          & $1.25\pm0.05$ & $1.20\pm0.05$ & $1.43\pm0.04$ & $1.31\pm0.03$ & $1.20\pm0.04$ & $1.16\pm0.04$ & $1.27\pm0.03$ & $1.21\pm0.03$  \\ 
$M_\star$                   &  $0.88\pm0.03$ & $0.66\pm0.10$ & $1.23\pm0.05$ & $2.01\pm0.17$ & $0.61\pm0.03$ & $0.76\pm0.04$ & $0.71\pm0.05$ & $0.97\pm0.08$  \\ 
$M^2_\star$                  & $0.70\pm0.09$ & $0.34\pm0.12$ & $1.17\pm0.16$ & $2.95\pm0.45$ & $0.24\pm0.13$ & $0.51\pm0.06$ & $0.32\pm0.19$ & $0.76\pm0.14$  \\ 
$M^3_\star$                  & $0.67\pm0.17$ & $0.18\pm0.10$ & $1.38\pm0.35$ & $4.3\pm1.0  $ & $0.11\pm0.26$ & $0.34\pm0.06$ & $0.16\pm0.40$ & $0.60\pm0.15$  \\ 
$1/r$                        & $1.47\pm0.05$ & $1.31\pm0.05$ & $1.71\pm0.03$ & $1.30\pm0.02$ & $1.25\pm0.04$ & $1.17\pm0.04$ & $1.40\pm0.02$ & $1.27\pm0.02$  \\ 
$z/r$                        & $1.52\pm0.06$ & $1.17\pm0.05$ & $1.90\pm0.04$ & $1.26\pm0.05$ & $1.30\pm0.04$ & $1.22\pm0.05$ & $1.47\pm0.06$ & $1.33\pm0.03$  \\ 
$M_\star/r$                  & $1.25\pm0.04$ & $0.61\pm0.05$ & $1.77\pm0.06$ & $2.03\pm0.19$ & $0.74\pm0.03$ & $0.86\pm0.05$ & $0.92\pm0.04$ & $1.06\pm0.07$  \\ 
$M^2_\star/r$                & $0.76\pm0.07$ & $0.35\pm0.06$ & $1.28\pm0.12$ & $3.1\pm0.7 $ & $0.28\pm0.08$ & $0.58\pm0.05$ & $0.38\pm0.11$ & $0.76\pm0.11$  \\ 
$M^3_\star/r$                & $0.56\pm0.11$ & $0.16\pm0.08$ & $1.15\pm0.20$ & $4.7\pm1.6  $ & $0.11\pm0.15$ & $0.36\pm0.06$ & $0.17\pm0.24$ & $0.57\pm0.11$  \\ 
$M^2_{\star,\mathrm{rms}}$   & $0.84\pm0.05$ & $0.59\pm0.09$ & $1.08\pm0.07$ & $1.72\pm0.14$ & $0.49\pm0.11$ & $0.71\pm0.04$ & $0.57\pm0.13$ & $0.87\pm0.08$  \\ 
$M^3_{\star,\mathrm{rms}}$   & $0.87\pm0.07$ & $0.56\pm0.09$ & $1.11\pm0.09$ & $1.62\pm0.13$ & $0.48\pm0.18$ & $0.70\pm0.04$ & $0.55\pm0.22$ & $0.84\pm0.07$  \\ 
$M^2_\star/r_\mathrm{,rms}$ & $0.87\pm0.04$ & $0.59\pm0.05$ & $1.13\pm0.05$ & $1.75\pm0.20$ & $0.53\pm0.06$ & $0.76\pm0.04$ & $0.62\pm0.08$ & $0.88\pm0.06$  \\ 
$M^3_\star/r_\mathrm{,rms}$ & $0.82\pm0.05$ & $0.54\pm0.07$ & $1.05\pm0.07$ & $1.68\pm0.18$ & $0.48\pm0.13$ & $0.71\pm0.04$ & $0.56\pm0.15$ & $0.83\pm0.05$  \\ 
$M_\star/r^3$                    & $3.8\pm0.2  $ & $0.56\pm0.08 $ & $6.2\pm0.3  $ & $1.8\pm0.3  $ & $3.25\pm0.16  $ & $0.82\pm0.09  $ & $5.05\pm0.25 $ & $1.31\pm0.13$  \\ 
$M_\star/r^2$                     & $2.2\pm0.1$ & $0.61\pm0.08$ & $3.25\pm0.15$ & $2.07\pm0.25      $ & $1.46\pm0.06$ & $0.87\pm0.05$ & $2.02\pm0.09$ & $1.21\pm0.08$  \\ 
$\sqrt{M_\star}/r$                        & $1.46\pm0.03$ & $0.89\pm0.06$ & $2.00\pm0.04$ & $1.68\pm0.10  $ & $1.05\pm0.02$ & $1.00\pm0.04$ & $1.26\pm0.02$ & $1.23\pm0.04$  \\ 
$\sqrt{M_h}/r$                    & $1.18\pm0.02$ & $1.01\pm0.05$ & $1.67\pm0.04$ & $1.42\pm0.07 $ & $0.76\pm0.04$ & $1.08\pm0.04$ & $0.92\pm0.06$ & $1.21\pm0.04$  \\ 
\hline
\end{tabular}
\\ 
{\footnotesize Medians of weighted galaxy counts for \hequad , inside various aperture radii and limiting magnitudes. The errors include, in quadrature, scatter from 10 samplings of redshift and stellar mass for each galaxy in \hequad , scatter from W1 and W3, BPZ - EAZY, and two different SExtractor detections.}
\label{tab:overdens}
\end{minipage}
\end{table*}

\begin{figure*}
\includegraphics[width=170mm]{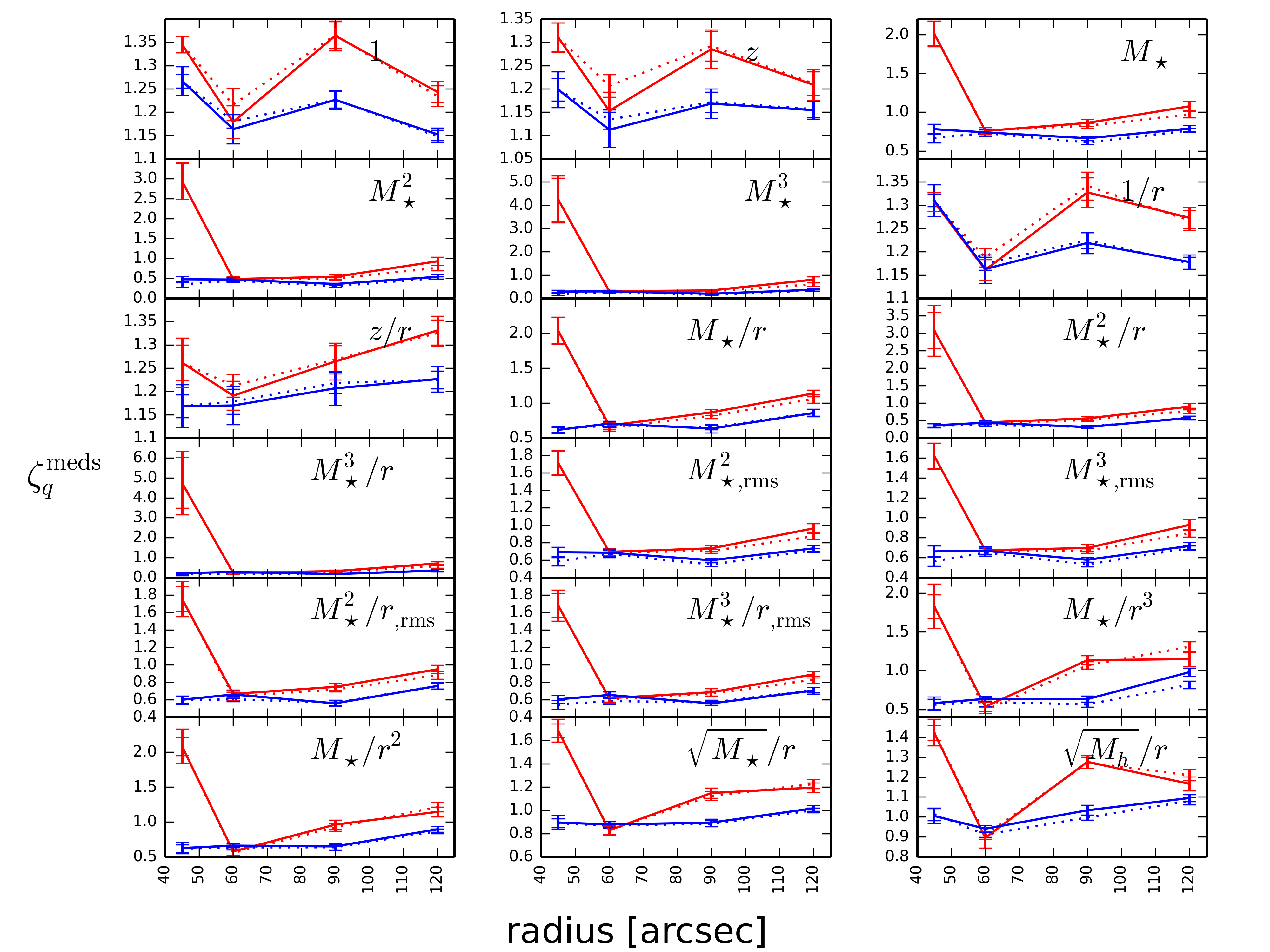}
\caption{Radial plot of the measured weighted count ratios $\zeta_q^\mathrm{meds}$, calculated for aperture radii of $45\arcsec$, $60\arcsec$, $90\arcsec$, and $120\arcsec$, using the combined CFHTLenS W1 and W3 fields. The blue line refers to $i\leq 24$, and the red line to $i\leq 23$. The solid line refers to redshifts estimated with BPZ, and the dotted line refers to redshifts determined with EAZY. The ranges of the vertical axes are different. Error bars include the scatter between W1-W4, and sampling from the galaxy magnitudes, redshifts and stellar masses (see text). They do not include scatter between different SExtractor parameters, which are included in Table \ref{tab:overdens}.
\label{fig:radial}}
\end{figure*}

Figure \ref{fig:radial} shows a radial plot of the measured overdensity for each weight, for four different aperture radii: $45\arcsec$, $60\arcsec$, $90\arcsec$ and $120\arcsec$. The \hequad\ field is comparatively more overdense for the brighter limiting magnitude ($i\leq23$) and, at the brighter limiting magnitude, for the $45\arcsec$ aperture. 

We note that the $1.27\pm0.05$ unweighted count overdensity we measure inside $45\arcsec$, for $i\leq24$, is larger that the underdensity of 0.89 ($\pm0.12$, assuming simple Poisson noise), measured by \citet{fassnacht11} inside the same aperture. This is likely due to the deeper magnitude limit they used, their much smaller control field, as well as possibly the use of a less careful masking technique. The present result supercedes the earlier analysis.

\subsection{Computing simulated $\zeta_q$ in the MS}\label{section:simulate}

Here, we compute weighted count ratios $\zeta_q$ from simulated fields obtained from the Millennium Simulation \citep[MS, ][]{springel05}, trying to closely reproduce the data quality of the \hequad\ and the CFHTLenS fields. We do this for two main reasons: First, since we will infer $P(\kext)$ by selecting lines of sight of specific overdensities from the MS, we need to ensure that it is fair to compare the overdensities in the MS to those in the real data. Second, by using the MS we can compare the overdensities we measure with their ``true'' values, and thus assess the quality of our estimates. 

The MS is an $N$-body simulation of cosmic structure formation in a cubic region $\sim680$ Mpc of co-moving size, with a halo mass resolution of $2 \times 10^{10}\,\mathrm{M}_{\odot}$ (corresponding approximately to a galaxy with luminosity $0.1L_\star$). Catalogues of galaxies populating the matter structures in the simulation were generated based on the semi-analytic galaxy models by \citet{delucia07}, \citet{guo11} and \citet[][]{bower06}. Furthermore, 64 simulated fields of $4\times4$ deg$^2$ where produced from the MS by ray-tracing \citep[][]{hilbert09}. These simulated fields contain, among other information, the observed positions, redshifts, stellar masses, and apparent magnitudes (e.g. in the SDSS $ugriz$ and 2MASS $JHK_s$ filters) of the galaxies in the field, as well as the gravitational lensing convergence $\kext$ and shear $\gext$ as a function of image position and source redshifts.

\begin{figure*}
\includegraphics[width=150mm]{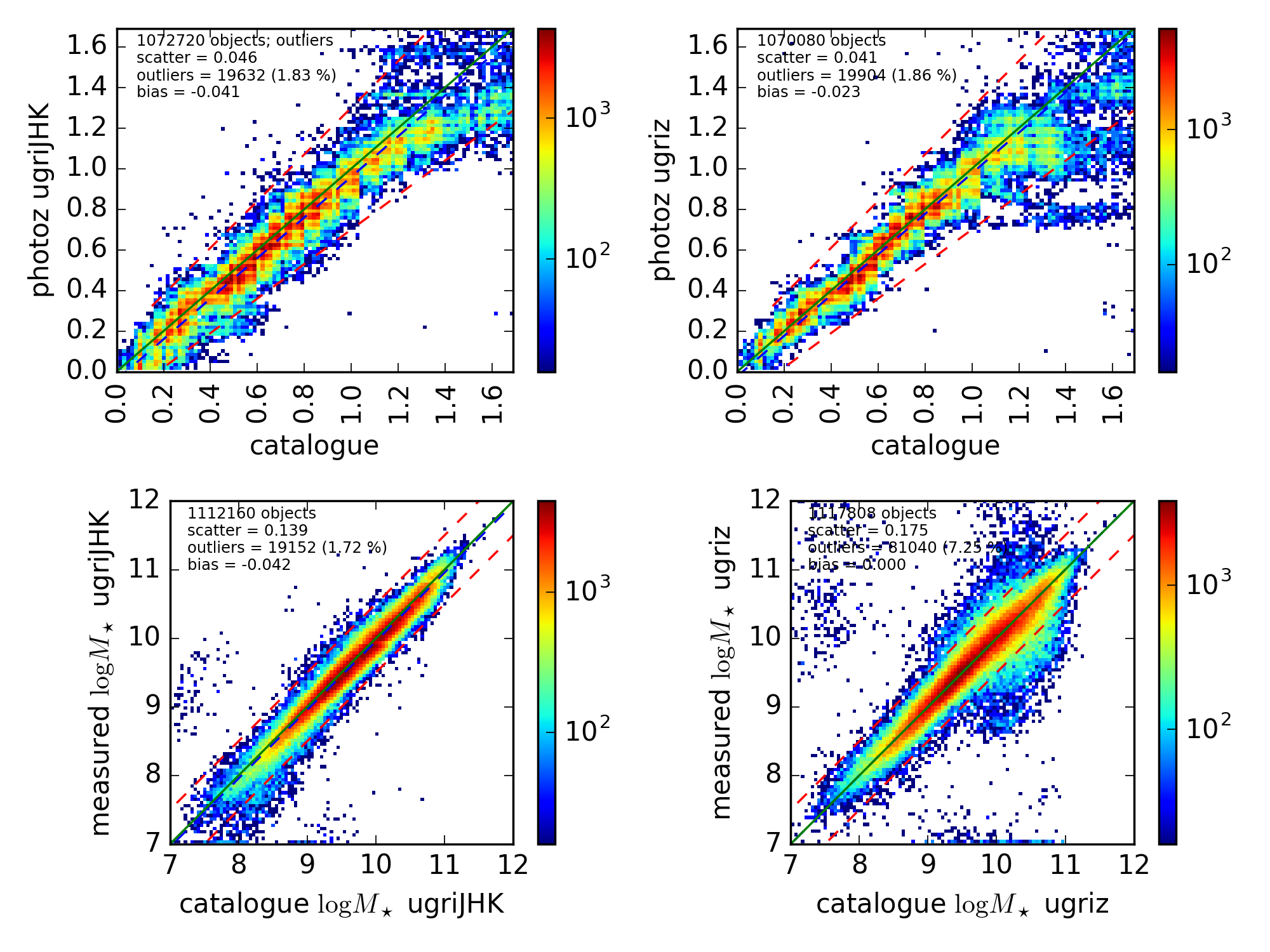}
\caption{Performance of the photometric redshift and stellar mass estimation in the MS, using the mock galaxy catalogue based on the \citet{delucia07} semi-analytic models, for galaxies inside a 4 deg $\times$ 4 deg field. Two different combinations of filters are used, as well as simulated photometric errors representative of the \hequad\ and CFHTLenS data. The bias for the photometric redshift when only $z<1$ objects are included decreases to -0.029 and -0.011 for the $ugriJHK_s$ and the $ugriz$ bands, respectively. For the lower plots, we define the outliers as $|\Delta \log M_\star| > 0.5$.
\label{fig:simphotozmstar}}
\end{figure*}

We use each of the MS fields, in turn, as fields whose overdensities
we want to measure (``\hequad -like fields''), as well as fields
against which we measure those overdensities (``control fields''). For
the \hequad -like fields we consider only their $ugriJHK_s$
photometry, whereas for the calibration fields we use their $ugriz$
photometry. Based on these, we compute photometric redshifts and
stellar masses for all $\sim70$ million $i<24$ mag galaxies, using the
same techniques we employed for the real data. This is because the
stellar masses and redshifts in our real data suffer from
observational uncertainties, which are not present in the available
synthetic catalogues. For each galaxy, we randomly sample its
``observed'' magnitude in a given band from a gaussian around its
catalogue magnitude, with a standard deviation given by the typical
photometric uncertainty of galaxies of similar magnitude in the real
data. In Figure~\ref{fig:simphotozmstar} we compare the redshifts and
stellar masses estimated for the galaxies in the MS with the catalogue
values, using photometry based on the \citet{delucia07} semi-analytic
models. We find better results compared to the catalogues based on
\citet{guo11} and \citet[][]{bower06}, and therefore we use the
\citet{delucia07}-based catalogue throughout this work. The
photometric redshift bias, scatter and fraction of outliers are
comparable to the ones measured for CFHTLenS and \hequad\ field
galaxies. We stress here that the superiority of the \citet{delucia07}
semi-analytic models is likely a consequence of these models being
more similar to the templates used by BPZ and Le PHARE. However, we
are only interested in the empirical result that by using these models
we obtain similar uncertainties in the simulations, and in the real
data. We thus conclude that we can indeed use the MS galaxy catalogue
to estimate overdensities with uncertainties similar to those found in
the real data.

We consider the same apertures and limiting magnitudes we used in the real data. In addition, we use the fact that a specific fraction of galaxies in the real \hequad \ field have spectroscopic redshift, as a function of magnitude and aperture radius. For these galaxies, we use their ``true'', catalogue redshifts. We calculate stellar masses with Le PHARE, in the same way we did for the real data, in particular using the same templates. There are, however, several differences to our approach, compared to the real data, which we present in Appendix~\ref{section:MSdetails}.

\begin{figure*}
\includegraphics[width=160mm]{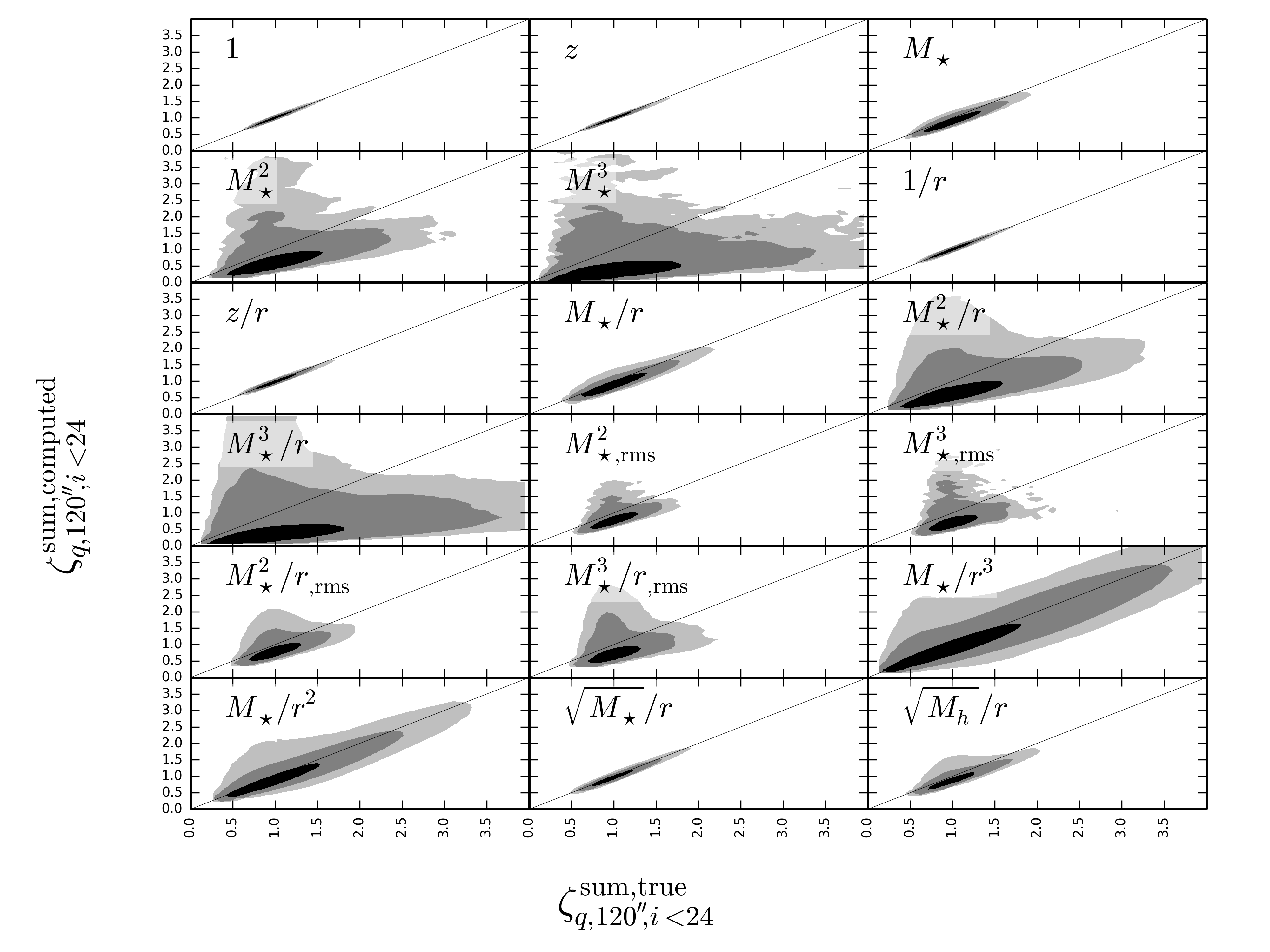}
\caption{Catalogue versus computed weighted ratios from the MS, using the mock galaxy catalogue based on the \citet{delucia07} semi-analytic models, for galaxies inside a 4 deg $\times$ 4 deg field. Each point represents $\zeta^{i\in\mathrm{MS},\mathrm{sum}}_q$ for $120\arcsec$-radius, $i\leq 24$ mag. Black, dark and light gray filled contours encompass regions of $1\sigma$, $2\sigma$, and $3\sigma$, respectively. The black line represents the diagonal.}
\label{fig:simsum}
\end{figure*}

\begin{figure*}
\includegraphics[width=160mm]{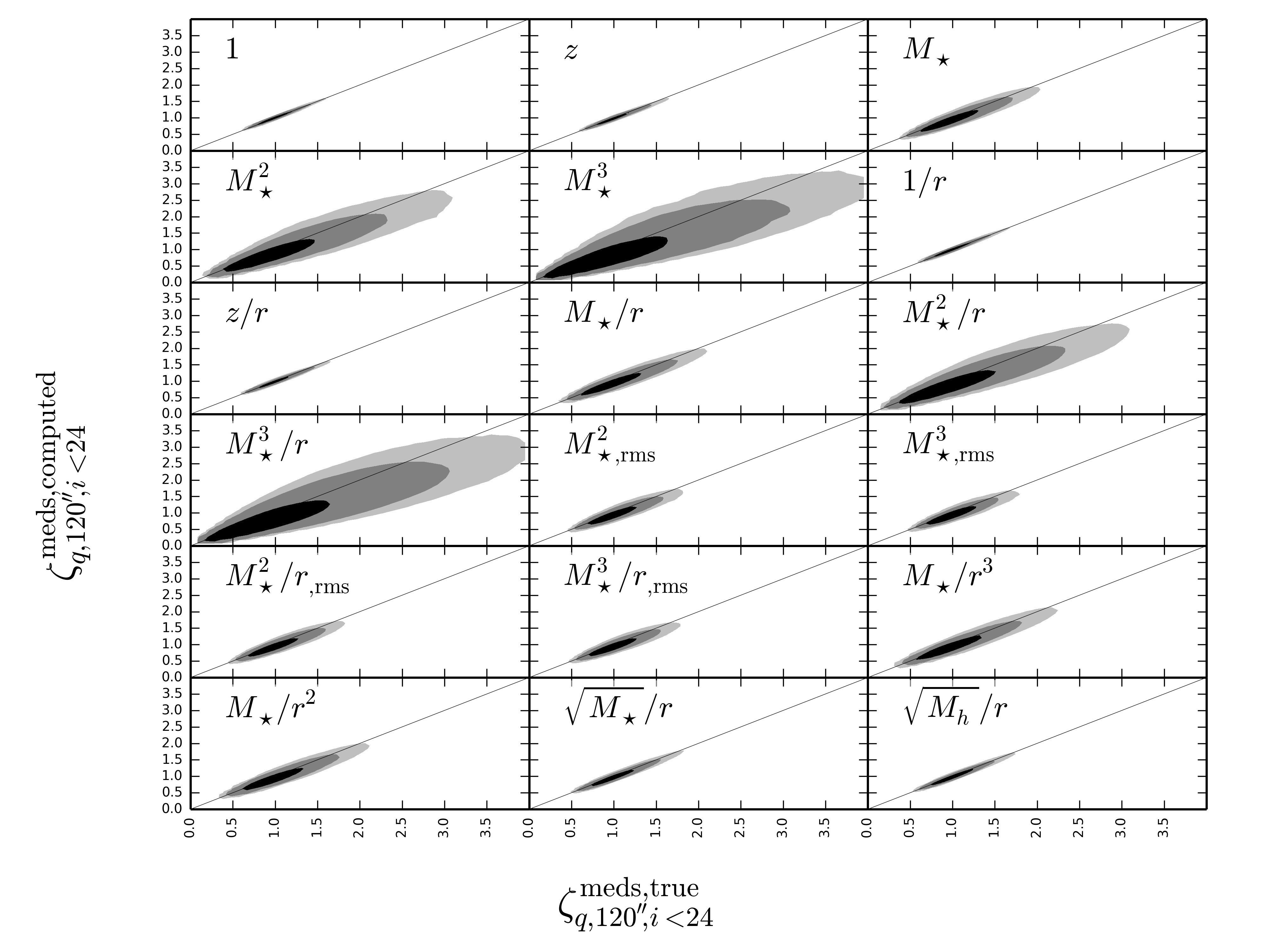}
\caption{Same as Figure \ref{fig:simsum}, but for $\zeta^{i\in\mathrm{MS},\mathrm{meds}}_q$.}
\label{fig:simmeds}
\end{figure*}

Next, we test how the ``measured'' overdensities compare to ``true'' overdensities, obtained by using the ``true'' values of redshift and stellar mass for each galaxy, readily available in the catalogue for the whole MS. We show the comparisons for $\zeta^{i\in\mathrm{MS},\mathrm{sum}}_q$ and $\zeta^{i\in\mathrm{MS},\mathrm{meds}}_q$ in Figures \ref{fig:simsum} and \ref{fig:simmeds}, respectively. $\zeta^{i\in\mathrm{MS},\mathrm{sum}}_q$ is a much noisier estimate than $\zeta^{i\in\mathrm{MS},\mathrm{meds}}_q$, and this is particularly obvious for all weights incorporating stellar mass, due to the high dynamic range of this quantity. This justifies our definition of $\zeta^{i\in\mathrm{MS},\mathrm{meds}}_q$ as a better estimate. 

We have also checked that a larger aperture radius and fainter magnitude limit produces smaller scatter, which is expected because they include more galaxies, resulting in less Poisson noise; the improvement is much more dependent on radius than on magnitude. 

\begin{figure*}
\includegraphics[width=130mm]{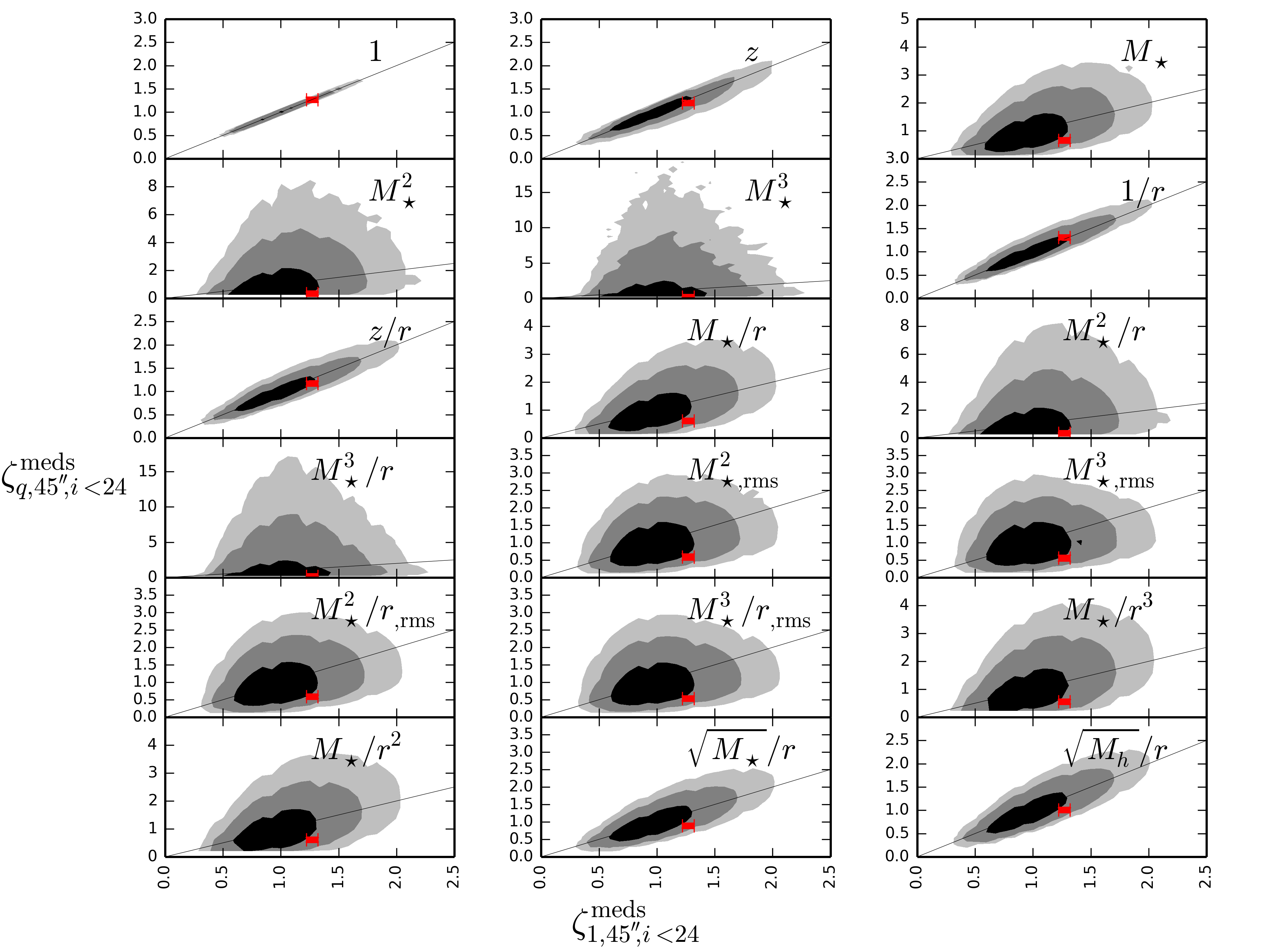}
\caption{Relation between number count ratios $\zeta_\mathrm{1}$ and weighted number count ratios $\zeta_q$, from the MS, using observational uncertainties similar to those of the \hequad\ field. The cells inside a $4\deg\times4\deg$ simulated field were used to construct the plot. The black, dark gray and light gray regions surround the 1-, 2-, and 3-$\sigma$ intervals, respectively. The black line represents the diagonal. The red error bars mark the measured overdensities for \hequad , and the associated 1-$\sigma$ error bars.}
\label{fig:galother}
\end{figure*}

Finally, in Figure \ref{fig:galother} we show the relations between the different $\zeta_q$. We find that the different $\zeta_q$ are correlated, as expected from their definitions, and that the specific values we determined for the \hequad \ field are realistic, in the sense that the they are expected at $\sim1$-$2\ \sigma$. We have checked that this result is robust to changing the aperture radius and limiting magnitude. 

By this point, we have related the $\kappa^i_\mathrm{ext}$ points (centers of each cell) in the 64 fields of the MS, where $i$ refers to each available cell, to their corresponding $\zeta^{i\in\mathrm{MS},\mathrm{meds}}_q$. In addition, we have also recorded the corresponding values of the shear $\gext^i$, to use as an additional constraint. H0LiCOW Paper IV measured a constant external shear strength (in addition to the shear stemming from explicit mass models of the strong-lens and nearby galaxies) which is close to the median of the shear distribution through the MS. This is helpful for ruling out high values from the $\kext$ distribution (see Figure 8 in \citet{collett16}.
Our use of all available $\kappa^i_\mathrm{ext}$ points in the MS (most of which are not strong-lensing lines of sight) is justified by \citet{hilbert09} and \citet{suyu10}, which showed that the distribution of $\kext$ from lines of sight to a strong lens is very similar to, and can be approximated by, the distribution for normal lines of sight (i.e. without a strong lens). We note that the redshift of the source quasar in \hequad , $z=1.69$, lies between two redshift planes in the MS, at $z=1.63$ and $z=1.77$. We therefore adopted the mean of the two planes for each value of the convergence $\kext$ and shear $\gext$. \footnotemark \footnotetext{While there are noticeable differences between individual values, we have determined $P(\kext)$ separately for a single plane, and found that the impact on the distribution is negligible, as the median of inferred P$(\kext)$ changes by only $\sim 0.002$ if we assume the source is at  $z=1.63$}.

\section{Determining $P(\kext)$}\label{section:kappa}

In the previous sections, we have explained how we estimate weighted count ratios for the real data, and analogously for the MS, and we have related every $\kext$ point in the MS to the corresponding weighted count ratio around its line of sight. We now present the mathematical formalism and implementation necessary to obtain the distribution of $\kext$ given our knowledge of weighted count ratios around \hequad .

\subsection{Theory and implementation}\label{section:theory}

We aim to estimate $P(\kext)$ using the MS catalogue of $\kext$ points, in a fully Bayesian framework. By $P(\kext)$ we refer to $p(\kext|\bold{d})$, where $\bold{d}$ stands for the available data, and we have made our dependence on the data explicit. The data refers to our catalogue of galaxies inside a given aperture and magnitude threshold, for both the \hequad \ and the CFHTLenS fields. It includes the galaxy number, galaxy positions in their respective apertures, as well as redshifts and stellar masses. In the sections above, we used these data in order to infer $\overline{\zeta^\mathrm{WX}_q}$, which we denote below as $\zeta_q$, and is by construction a noisy quantity. We use $\zeta_q$ as a random variable, whose connection to the data and the external convergence can be expressed by a joint distribution $p(\kext,\zeta_q,\bold{d})$. Then, $p(\kext|\bold{d})$ can be expressed as:

\begin{equation}
p(\kext|\bold{d}) = \frac{p(\kext,\bold{d})}{p(\bold{d})} = \int d \zeta_q \frac{p(\kext,\zeta_q,\bold{d})}{p(\bold{d})}         \ . 
\end{equation}

\noindent Next, we make the assumption that 

\begin{equation}
p(\bold{d}|\kext,\zeta_q) = p(\bold{d}|\zeta_q)        \ , 
\end{equation}

\noindent i.e. the likelihood of the data does not explicitly depend on the external convergence, for fixed $\zeta_q$. This is justified, since we have defined $\zeta_q$ based solely on the data, without reference to the external convergence. From this,

\begin{multline}
p(\kext,\zeta_q,\bold{d}) = p(\kext,\zeta_q) p(\bold{d}|\kext,\zeta_q) = \\ 
p(\kext,\zeta_q) p(\bold{d}|\zeta_q)  = p(\kext,\zeta_q) \frac{p(\zeta_q,\bold{d})}{p(\zeta_q)}   \ ,
\end{multline}

\noindent and thus 

\begin{multline}
p(\kext|\bold{d})  = \int d \zeta_q \frac{p(\kext,\zeta_q) p(\zeta_q,\bold{d})}{p(\zeta_q) p(\bold{d})}  = \\ 
\int d \zeta_q p(\kext|\zeta_q) p(\zeta_q|\bold{d})   \ .
\end{multline}

\noindent That is, given our estimate of $\zeta_q$ from the data, by using a correspondence between $\zeta_q$ and $\kext$, we obtain the $\kext$ distribution. Here, we consider $p(\zeta_q|\bold{d})\equiv N_q\left(\zeta_{q};\sigma_{\zeta_{q}}\right)$ to be a gaussian with mean and standard deviation given by Table \ref{tab:overdens}, and we make use of the MS by replacing $p(\kext|\zeta_q)$ with $p_\mathrm{MS}(\kext|\zeta^\mathrm{MS,meds}_q\equiv \zeta_q)$.

As mentioned in Section \ref{section:intro}, G13 showed that the standard deviation of $P(\zeta_q|\bold{d})$, which we denote as $\sigma_\kappa$, can decrease when information is added by using multiple conjoined weights. They found the best improvement when using combinations of three weights, including $q_\mathrm{gal}$ and $q_{1/r}$. We make use of this result, and consider a third weight from those in Section \ref{section:ratiosdescript}, in addition to the shear constraint. Thus, our distribution becomes

\begin{multline}
p(\kext|\bold{d}) = \int d \zeta_1 d \zeta_{1/r} d \zeta_{q\neq1,1/r} d \zeta_{\gext} p_\mathrm{MS}(\kext| \zeta_1, \zeta_{1/r}, ... \\  
... \zeta_{q\neq1,1/r}, \zeta_{\gext}) p(\zeta_1, \zeta_{1/r}, \zeta_{q\neq1,1/r}, \zeta_{\gext} | \bold{d})  \ .
\end{multline}

\noindent We determined $p(\zeta_q|\bold{d})$ from the data independently for each $q$, as gaussians much narrower than the distributions whose medians they represent (e.g., Figures \ref{fig:12024meds} and \ref{fig:4523meds}). We can thus factorize 

\begin{multline}
p(\zeta_1, \zeta_{1/r}, \zeta_{q\neq1,1/r}, \zeta_{\gext} | \bold{d}) \simeq \\ 
p(\zeta_1 | \bold{d}) p(\zeta_{1/r} | \bold{d}) p(\zeta_{q\neq1,1/r} | \bold{d}) p(\zeta_{\gext}) | \bold{d}).
\label{eq:approx} 
\end{multline}
 
 \noindent We remind the reader that in general (i.e. over the whole extent of their distribution) the $\zeta_q$ are correlated, as we have seen in Section \ref{section:simulate}, and not independent. \footnotemark \footnotetext{We tested that the approximation in Equation \ref{eq:approx} is justified, by measuring the correlation coefficients between $\zeta_1$, $\zeta_{1/r}$, and $\zeta_{q\neq1,1/r}$ to be $\sim 0$ (at most $\sim0.2$, in rare cases), for the relevant narrow range of interest.} 
 
G13 showed that simply adding up $\kext$ points corresponding to lines of sight with $N_\mathrm{gal}\in\zeta_{q_\mathrm{gal}}\overline{N_\mathrm{gal}} \pm \mathrm{E}_{q_\mathrm{gal}}$ (this generalizes to $(W_q/\overline{W_q})\overline{N_\mathrm{gal}}\in\zeta_q\overline{N_\mathrm{gal}} \pm \mathrm{E}_q$), would bias $P(\kext)$. Here $\overline{N_\mathrm{gal}}$ is the median number of galaxies in an aperture of interest around a given line of sight from the MS, and $\mathrm{E}_q$ we choose to be twice the width of $p(\zeta_q|\bold{d})$. The bias comes from the fact that, e.g., for a relatively overdense field, the number of lines of sight $N_\mathrm{LOS}$ available with a galaxy count $N_\mathrm{gal}$ will be larger than that with a galaxy count $N_\mathrm{gal}+1$ (i.e., there are comparatively fewer fields more overdense than a field which is already overdense). A larger number of lines of sight means that their respective $\kext$ distribution will be overrepresented, and the overall $P(\kext)$ will be biased towards those values. The solution adopted by G13 is to divide the $2\mathrm{E}_q$ interval into $2\mathrm{E}_q$ bins of individual length 1\footnotemark \footnotetext{In practice, in order to reduce dimensionality, we allow the bins to be as large as 2. G13 (see their Figure 1) showed that this introduces negligible differences.} 
 (for $\zeta_{q_\mathrm{gal}}=1$ this corresponds to incrementing $\overline{N_\mathrm{gal}}$ by 1), and weight the $\kext$ distribution in each of the bins by $1/N_\mathrm{LOS}$, where $N_\mathrm{LOS}$ is the number of lines of sight in that particular bin. This way, each of the $2\mathrm{E}_q$ $\kext$ distributions carries equal weight into the combined distribution. In our case, we typically use four conjoined constraints $\{q_i,q_j,q_k,q_l\}=\{q_\mathrm{gal},q_{1/r},q\neq\{1,1/r\},q_{\gext}\}$, and therefore have $2\mathrm{E}_{q_i} \cdot 2\mathrm{E}_{q_j} \cdot 2\mathrm{E}_{q_h} \cdot 2\mathrm{E}_{q_k}$ multidimensional bins. 
 
 We account for the bias discussed above and compute $p(\kext|\bold{d})$ as a series of nested sums 
 
 \begin{multline}
\sum_{i \in}^{\zeta_{q_i}\overline{N_\mathrm{gal}} \pm \mathrm{E}_i} \sum_{j \in}^{\zeta_{q_j}\overline{N_\mathrm{gal}} \pm \mathrm{E}_j} \sum_{k \in}^{\zeta_{q_k}\overline{N_\mathrm{gal}} \pm \mathrm{E}_k}  \sum_{l \in}^{\zeta_{q_l}\overline{N_\mathrm{gal}} \pm \mathrm{E}_l}                     p_\mathrm{MS}(\kext| \zeta_{q_i}, ... \\ 
... \zeta_{q_j}, \zeta_{q_k}, \zeta_{q_l}) \frac{\prod_{x={i,j,k,l}} N_x\left(\zeta_{q_x};\sigma_{\zeta_{q_x}}\right)}{N^{(i,j,k,l)}_\mathrm{LOS}}   
\label{eq:p}              
\end{multline}
 
\noindent where  and $N^{(i,j,k,l)}_\mathrm{LOS}$ is the number of lines of sight in each multidimensional bin with indices $(i,j,k,l)$, and $p_\mathrm{MS}$ is the distribution of $\kext$ corresponding to each of these lines of sight.
 
 For brevity, we refer to $p(\kext|\bold{d})$ implemented by Equation \ref{eq:p} as $P(\kext|\zeta_1,\zeta_{1/r},\zeta_{q\neq1,1/r},\zeta_{\gext})$. We also consider selected distributions with fewer constraints. There are two practical limitation in not using more than four conjoined constraints. First, applying Equation \ref{eq:p} is computationally intensive, and scales quickly with the number of dimensions. Second, the MS contains a limited number of $\kappa$ points, and the number of such points included in a bin decreases as additional constraints are added. 

\subsection{Testing for biases using simulated data}\label{section:biases}

\begin{figure*}
\includegraphics[width=175mm]{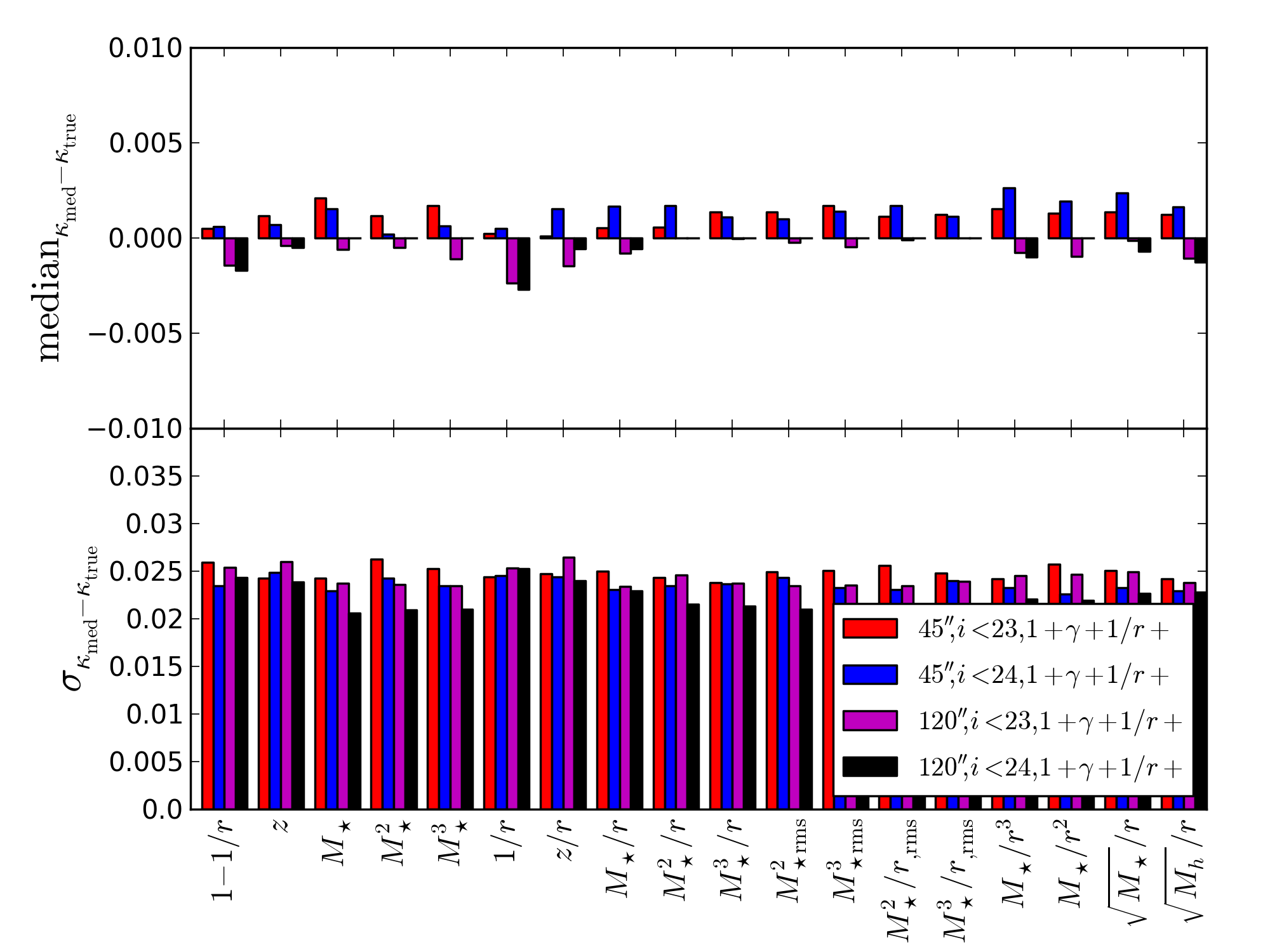}
\caption{ Medians and standard deviations of the $\kappa - \kappa_\mathrm{med}$ distributions for a variety of aperture radii, limiting magnitudes and conjoined weights ($1,\gext,1/r,+$). Each point in the distribution represents one of 5000 cells from the MS, which are similar in overdensity to \hequad .}
\label{fig:kappacomp}
\end{figure*}

\begin{figure}
\includegraphics[width=85mm]{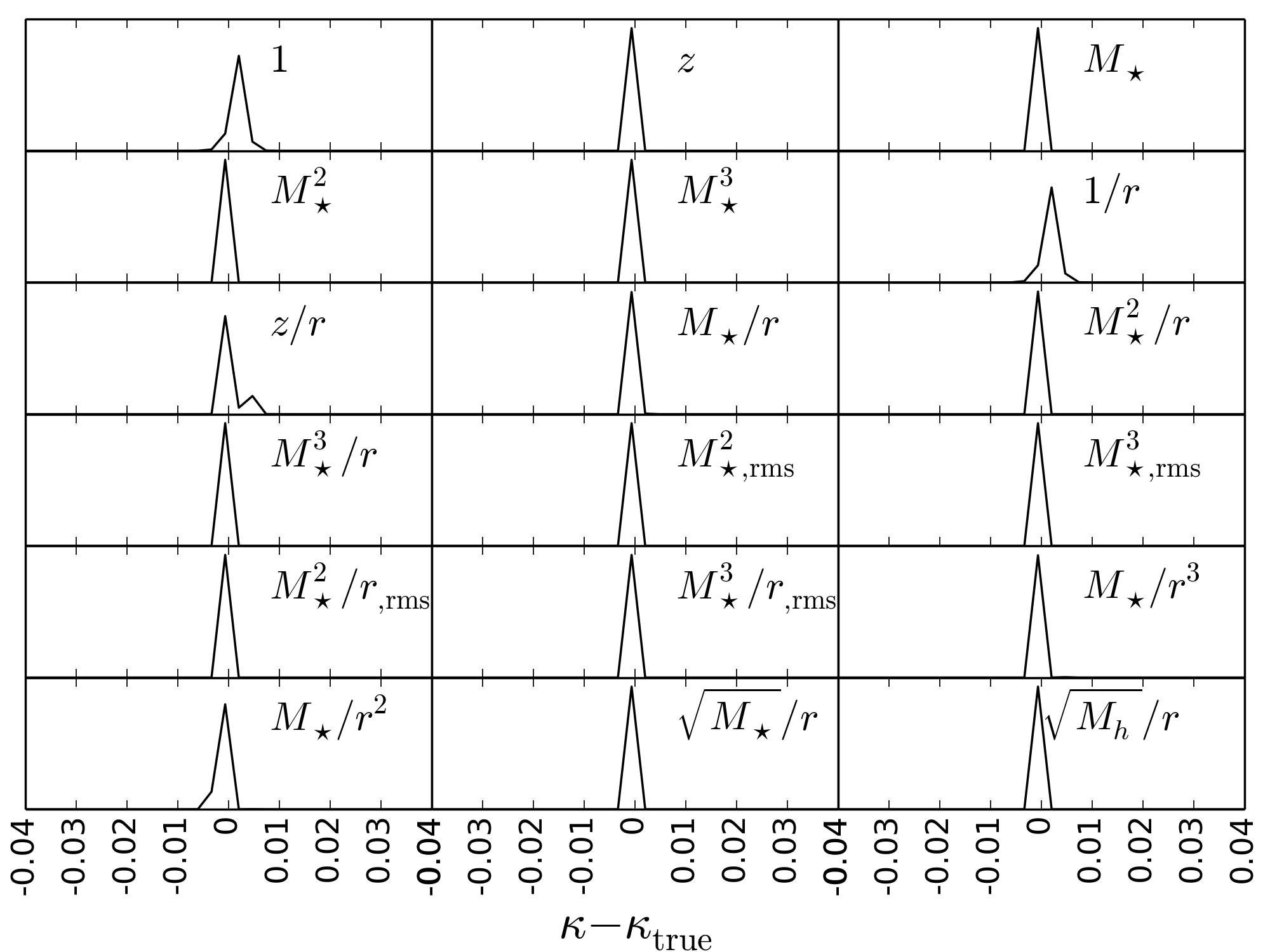}
\caption{Products of 100 $P(\kappa - \kappa_\mathrm{med})$ distributions, computed in a similar way to that of \hequad\ for cells of similar overdensities, using as constraints $1+\gext+1/r\ +$ one other weight, within a $120\arcsec$ aperture, $i<24$ mag. The plots for other apertures and magnitude limits are similar.}
\label{fig:kappaphil}
\end{figure}

It is possible to use the MS itself to estimate the accuracy of our $p(\kext|\bold{d})$ estimation, and test for biases. First, we randomly select 5000 cells from the MS, which are similar in terms of overdensity to \hequad . We then estimate $p(\kext|\bold{d})$ for each of them. However, since this estimation would be computationally expensive, we consider very small uncertainties around the computed overdensities, so that Equation \ref{eq:p} reduces to the computation of a single distribution, in one bin. For each of the 5000 distributions, we record its median, $\kappa^\mathrm{med}_\mathrm{ext}$. We then determine the distribution of $\kappa^\mathrm{med}_\mathrm{ext} - \kappa^\mathrm{true}_\mathrm{ext}$, where $\kappa^\mathrm{true}_\mathrm{ext}$ is the true value at the center of each cell. We plot in Figure \ref{fig:kappacomp} the median and standard deviation of this distribution, for each weight combination, as well as aperture radius and limiting magnitude. We find that $\kappa^\mathrm{med}_\mathrm{ext}$ is typically an unbiased estimate of $\kappa^\mathrm{true}_\mathrm{ext}$, to better than $\lesssim0.0025$. For the $45\arcsec$ aperture $\kappa^\mathrm{med}_\mathrm{ext}$ seems to slightly overestimate $\kappa^\mathrm{true}_\mathrm{ext}$, whereas the $45\arcsec$ aperture shows the opposite tendency. These estimates are noisy, with a standard deviation of $\sim0.020-0.025$. This is to be expected: being the median of a distribution of $\kappa^\mathrm{true}_\mathrm{ext}$ values, $\kappa^\mathrm{med}_\mathrm{ext}$ cannot vary too much, compared to the individual $\kappa^\mathrm{true}_\mathrm{ext}$ points. However, the standard deviations of the 5000 individual distributions are also $\sim0.025$, which means that $\kappa^\mathrm{true}_\mathrm{ext}$ is typically well-contained inside the individual distributions. 

Next, we follow the example of \citet{collett13} in assessing the presence of biases in our estimation of the full $p(\kext|\bold{d})$ distribution. In the absence of biases, $p(\kext-\kappa^\mathrm{true}_\mathrm{ext}|\bold{d})$ is centered on zero. For different cells, these offset distributions can be multiplied together, resulting in a narrower distribution $P_N=\prod_{i=1}^N p_i(\kext-\kappa^\mathrm{true}_\mathrm{ext}|\bold{d})$. Offsets from zero in the centroid of this distribution would be indicative of biases. We show the results of this approach in Figure \ref{fig:kappaphil}, where we adopt $N=100$, and find no indication of offsets for any of the weights we consider. We conclude that, for fields of overdensity similar to \hequad , our technique is not affected by biases. 

\section{Results and Discussion}\label{section:discuss}

\begin{figure}
\includegraphics[width=90mm]{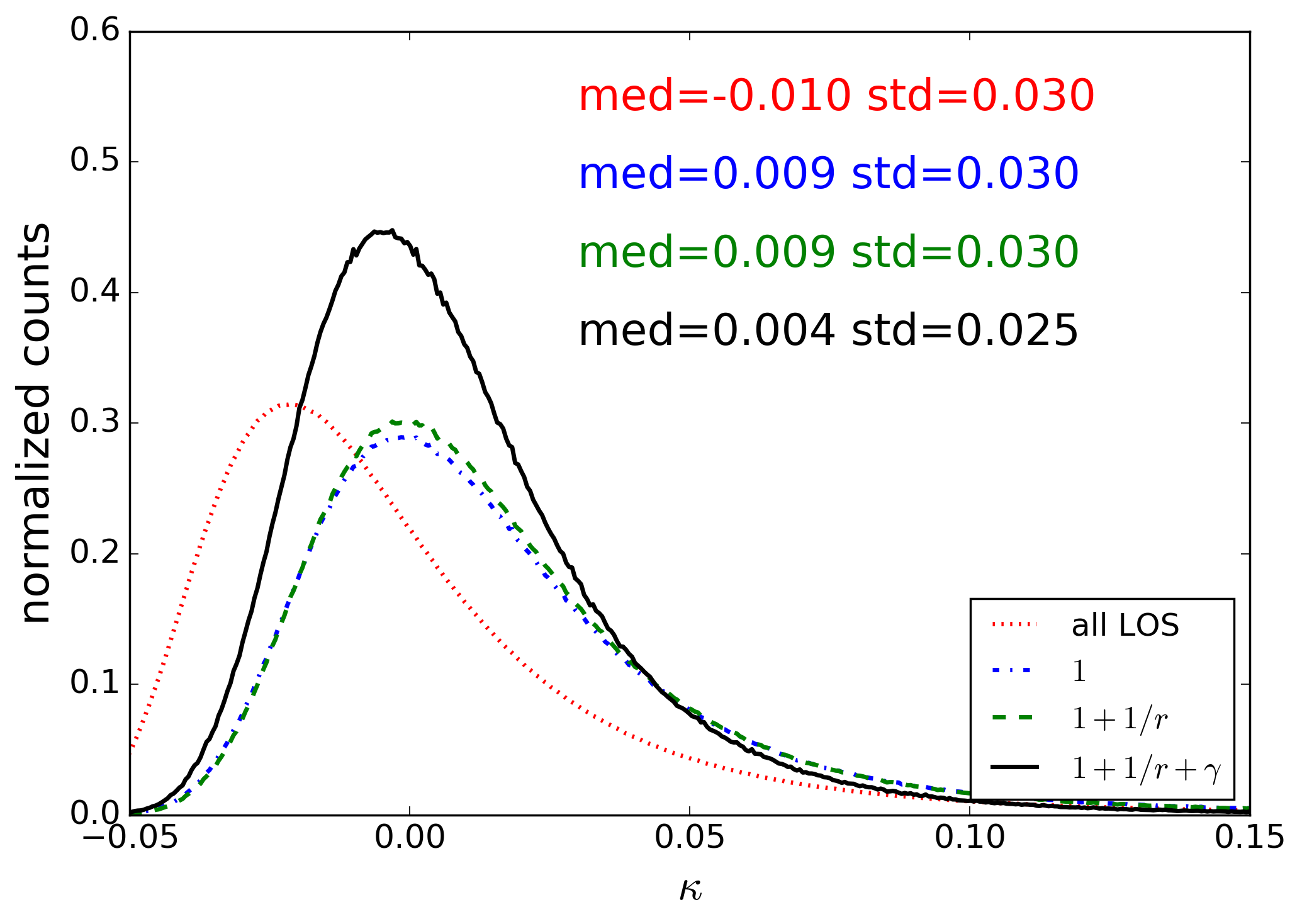}
\caption{Example of the variation of $P(\kext)$ with the addition of constraints, for aperture radius $45\arcsec$, $i<24$ mag.}
\label{fig:kappaselect}
\end{figure}

\begin{table} 
\tiny
 \centering
 \begin{minipage}{155mm}
  \caption{$\kext^\mathrm{med}$ and $\sigma_\kappa$ for conjoined weights $1+\gext+1/r+q$}
  \begin{tabular}{@{}lllll@{}}
  \hline 
& $45\arcsec$ & $45\arcsec$ & $120\arcsec$ & $120\arcsec$ \\
q & $i < 24$ & $i < 23$ & $i < 24$ & $i < 23$ \\
 \hline
$1-\frac{1}{r}$               			& $+0.002, 0.025$ & $-0.001, 0.025$ & $+0.002, 0.024$ & $+0.002, 0.025$   \\ 
$z$                          & $+0.003, 0.025$ & $-0.004, 0.024$ & $+0.001, 0.025$ & $+0.002, 0.025$   \\ 
$M_\star$                   &  $-0.001, 0.023$ & $+0.001, 0.025$ & $-0.002, 0.023$ & $-0.000, 0.024$   \\ 
$M^2_\star$                  & $-0.002, 0.023$ & $+0.002, 0.025$ & $-0.002, 0.023$ & $-0.000, 0.024$   \\ 
$M^3_\star$                  & $-0.001, 0.023$ & $+0.002, 0.025$ & $-0.002, 0.023$ & $-0.000, 0.024$   \\ 
$\frac{1}{r}$                        & $+0.004, 0.025$ & $-0.003, 0.024$ & $+0.000, 0.025$ & $+0.002, 0.025$   \\ 
$\frac{z}{r}$                        & $+0.003, 0.025$ & $-0.004, 0.024$ & $+0.004, 0.026$ & $+0.007, 0.027$   \\ 
$\frac{M_\star}{r}$                  & $-0.002, 0.023$ & $+0.002, 0.025$ & $-0.002, 0.023$ & $+0.000, 0.024$   \\ 
$\frac{M^2_\star}{r}$                & $-0.002, 0.023$ & $+0.002, 0.025$ & $-0.002, 0.023$ & $-0.001, 0.024$   \\ 
$\frac{M^3_\star}{r}$                & $-0.002, 0.023$ & $+0.002, 0.025$ & $-0.002, 0.023$ & $-0.001, 0.024$   \\ 
$M^2_{\star,\mathrm{rms}}$   & $-0.002, 0.023$ & $+0.002, 0.025$ & $-0.002, 0.023$ & $-0.000, 0.024$  \\ 
$M^3_{\star,\mathrm{rms}}$   & $-0.002, 0.023$ & $+0.001, 0.025$ & $-0.002, 0.023$ & $-0.001, 0.024$   \\ 
$\frac{M^2_\star}{r}_\mathrm{,rms}$ & $-0.002, 0.023$ & $+0.002, 0.025$ & $-0.002, 0.023$ & $-0.001, 0.024$   \\ 
$\frac{M^3_\star}{r}_\mathrm{,rms}$ & $-0.002, 0.023$ & $+0.002, 0.025$ & $-0.002, 0.023$ & $-0.001, 0.024$  \\ 
$\frac{M_\star}{r^3}$                   & $-0.002, 0.023$ & $+0.001, 0.025$ & $-0.002, 0.023$ & $+0.002, 0.025$  \\ 
$\frac{M_\star}{r^2}$                      & $-0.002, 0.023$ & $+0.002, 0.025$ & $-0.002, 0.023$ & $+0.002, 0.025$   \\ 
$\frac{\sqrt{M_\star}}{r}$                        & $-0.002, 0.023$ & $+0.002, 0.025$ & $-0.002, 0.023$ & $+0.002, 0.025$   \\ 
$\frac{\sqrt{M_h}}{r}$                    & $-0.001, 0.023$ & $-0.001, 0.024$ & $-0.001, 0.024$ & $+0.001, 0.025$  \\ 
\hline
\end{tabular}
\\ 
{\footnotesize The pairs on each column represent ($\kext^\mathrm{med}$, $\sigma_\kappa$). Here $q=1-1/r$ \\  refers to conjoined weights $1+\gext$. }
\label{tab:kappa}
\end{minipage}
\end{table}

\begin{figure*}
\includegraphics[width=175mm]{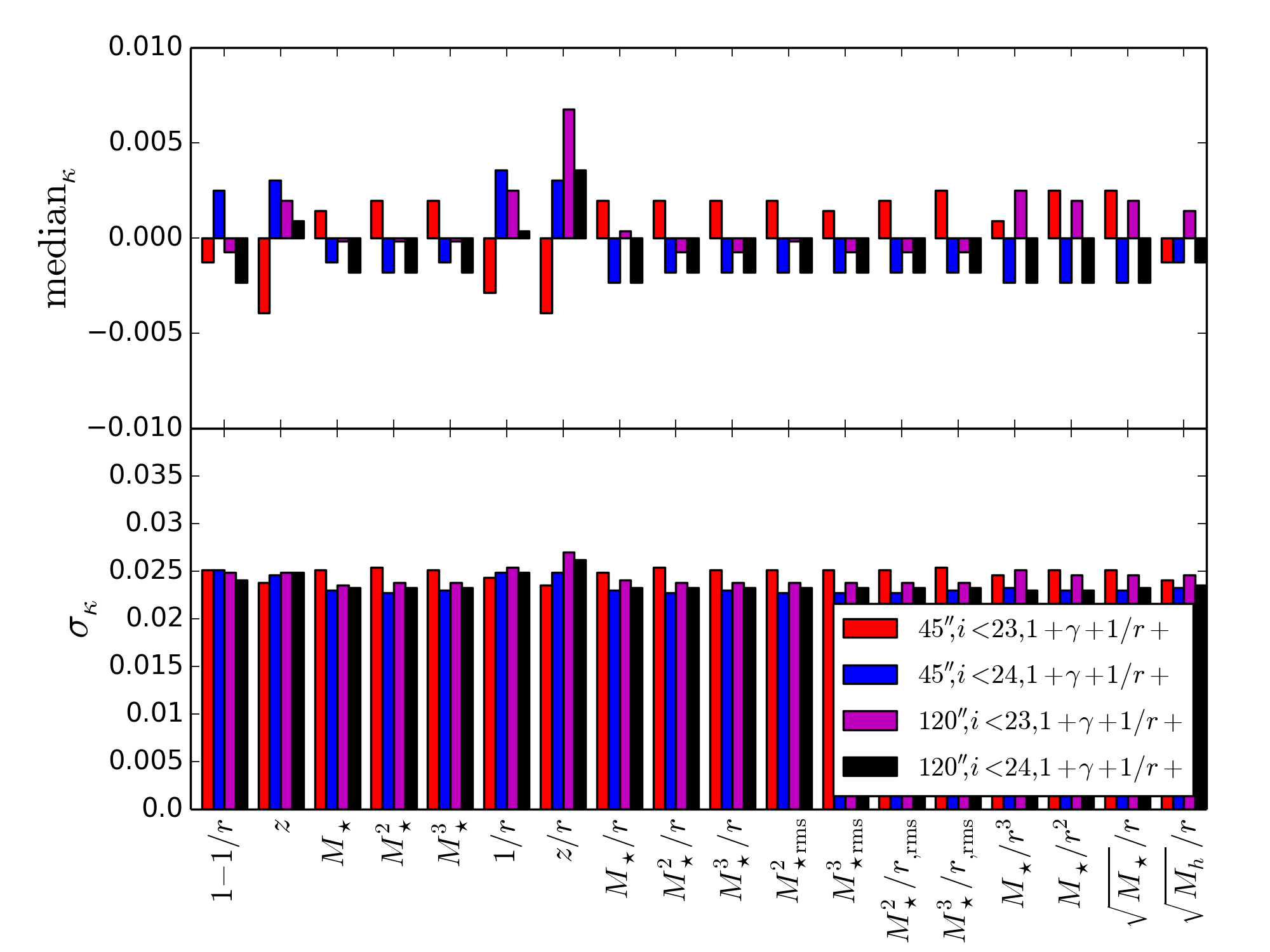}
\caption{ Medians and standard deviations of the $P(\kappa|1,\gext,1/r,+)$ distributions for a variety of aperture radii, limiting magnitudes and conjoined weights, for \hequad . Here $1-1/r$ refers to $P(\kappa|1,\gext)$.
\label{fig:kappa}}
\end{figure*}

We first present the results on the distribution of external convergence in Figure \ref{fig:kappaselect}. The \hequad \ field is slightly overdense in terms of unweighted galaxy counts for aperture radius $45\arcsec$, $i<24$ mag, $P(\kext|\zeta_{q_\mathrm{gal}})$ resulting in a slightly positive $\kappa_\mathrm{ext}^\mathrm{med}$ of 0.009. The addition of the radial dependence constraint, $P(\kext|\zeta_{q_\mathrm{gal}},\zeta_{q_{1/r}})$, has a very small effect on the distribution. As expected, since the measured shear is similar to the median one through the MS, adding the shear constraint $P(\kext|\zeta_{q_\mathrm{gal}},\zeta_{q_{1/r}},\zeta_{\gext})$ has the effect of narrowing the distribution, and moving it towards lower $\kappa_\mathrm{ext}^\mathrm{med}$ of 0.004. 

We show the resulting medians and standard deviations of the distributions for all weight combinations, as well as aperture radii and limiting magnitudes, in Figure \ref{fig:kappa}, and summarize the results in Table \ref{tab:kappa}. We find that the addition of weighted count constraints, on top of the constraints from shear, unweighted number counts and distance to the lens, only moves the peak and width of the distributions by $\sim0.005$. This is expected, since G13 find that the use of weighted count constraints does not yield much improvement for fields of typical overdensities, such as \hequad . As a result, we do not expect further improvement if using more than 4 conjoined constraints. The standard deviations of each of the distributions are $\sim 0.025$, which is similar to the values G13 find for fields of comparable overdensities; we note, however, that G13 did not use shear as a constraint. \footnotemark \footnotetext{We also checked that changing the shear constraint by $\sim 0.025$ towards lower values lowers $\kext$ by $\sim0.005$.}

The shift value at which the distributions are consistent with each other, $\sim0.005$, even if different apertures and limiting magnitudes are considered, corresponds to $\sim 0.5\%$ impact on $H_0$, according to Equation \ref{eq:H}. Combined with the result from Section \ref{section:biases}, that our technique is free of biases, this means that our approach is insensitive to the exact choice of aperture and limiting magnitude, among those we explored, at this level. That is, our small, bright limit is already large and deep enough for our analysis. In contrast, as we consider larger and larger apertures, we would expect that we wash away signal, unless we weigh by something steeper than 1/r, because we include larger numbers of galaxies which may be too distant to contribute to $\kappa^\mathrm{true}_\mathrm{ext}$. Given large enough apertures, they will tend to an unweighted count ratio of unity regardless of the field. The same argument would hold for deeper magnitudes, except that we implement a cut at the redshift of the source quasar, so going deeper does not imply that we contaminate the signal. The consistency of our results indicates that our large, deep limits are still sensitive to the desired $\kappa^\mathrm{true}_\mathrm{ext}$. Finally, the mutual consistency of the distributions for the two limiting magnitudes also ensures that our results are not affected by possible incompleteness\footnotemark \footnotetext{Though an estimate of completeness is not available for CFHTLenS, for the shallower CFHTLS parent catalogue this is 80\% for extended sources of $i \sim 23.4$ mag, according to \url{http://www.cfht.hawaii.edu/Science/CFHLS/T0007/T0007-docsu12.html}}. 

We note that the small $\kext$ value we measure, well consistent with zero, is also in agreement with the weak lensing upper limit on convergence $<0.04$ for this system (Tihhonova et al., in prep.), and the unlikely existence of large structures such as groups, significant enough to boost the convergence (H0LiCOW Paper II).

\section{Conclusions and future work}\label{section:concl}

In this work, we aimed to estimate a robust probability distribution function of the external convergence for \hequad , in order to enable the use of this lens system as an accurate probe of $H_0$. We used spectroscopy and multiband images of the \hequad \ field, and we used the wide component of CFHTLenS as a control field. Building on the work by G13, we refined the method in order to cope with the large fraction of masks in our control field, and we also used more robust medians rather than sums in order to compare weighted counts. We thoroughly explored sources of error in our data sets, such as mask coverage, galaxy-star classification, detection efficiency etc.; we propagated these into the computation of weighted count ratios, finding that the \hequad \ field is more overdense, in terms of number counts, than previously estimated. We used the whole extent of the MS to simulate photometric data of the same quality, and connect the MS lensing convergence catalogue to synthetic weighted count ratios estimated in a similar way. We than estimated the probability distribution function of the external convergence for fields similar in overdensity to \hequad, in a Bayesian, unbiased way.

We considered multiple aperture radii and limiting magnitudes, and tested them using the MS, finding that a $45\arcsec$ aperture and a limiting magnitude of $i\leq23$ provide enough spatial coverage and depth to estimate the distribution of external convergence via the weighted counts technique. We find that our different estimates are consistent with each other at a level of $\sim 0.005$, corresponding to $\sim 0.5\%$ impact on $H_0$. Our estimate which is least affected by photometric redshifts and stellar mass uncertainties, $P(\kext|\zeta_{q_\mathrm{gal}},\zeta_{q_{1/r}},\zeta_{\gext})$, has a median of 0.004, and a standard deviation of 0.025. This uncertainty contributes $\sim2.5\%$ rms error to the value of $H_0$. We intend to employ the techniques developed in this paper for the analysis of the other H0LiCOW lens systems. In particular, \hequad\ is a rather typical line of sight, and we expect that lenses residing in comparatively overdense fields will benefit more from the use of additional constraints including photometric redshifts and stellar masses.

Throughout this work, we have made extensive use of the MS. The weighted count ratios technique is designed to minimize our reliance on a particular simulation, but it will be useful to repeat this analysis by using simulations for different cosmologies and galaxy models to test any remaining dependencies. However, we expect such dependencies to be small, given that the external convergence we measure is close to zero. Assuming a simple linear deterministic galaxy bias model, the convergence inferred from a given relative galaxy number overdensity scales roughly with the mean matter density parameter $\Omega_\mathrm{m}$ and the matter density fluctuation amplitude $\sigma_8$ (see Section \ref{section:cosm_depend}). Therefore, for example, $\kappa^\mathrm{med, Planck}_\mathrm{ext} \propto \kappa^\mathrm{med,MS}_\mathrm{ext}\Omega_\mathrm{m}^\mathrm{Planck}\sigma_8^\mathrm{Planck}/(\Omega_\mathrm{m}^\mathrm{MS}\sigma_8^\mathrm{MS}) \sim 1.13\kappa^\mathrm{med,MS}_\mathrm{ext}$. For $\kappa^\mathrm{med,MS}_\mathrm{ext}=0.004$, this corresponds to $\lesssim 0.001$ impact. We leave further checks for future work, as other simulations with convergence maps become available.

Recently, \citet{mccully16} presented a technique of reconstructing the external convergence without relying on a particular simulation, through a direct modelling of the field. This has the potential of further reducing the uncertainty on the external convergence. This work has produced the galaxy catalogues necessary for a future implementation of that technique. While we have accounted in this work for the presence of voids, groups and clusters statistically, through the use of the MS, our catalogue products are also used in separate works (H0LiCOW Paper II and Tihhonova et al., in prep.) to directly identify such structures. 

\section*{Acknowledgments}

The authors would like to thank Jean Coupon, Thomas Erben, Hendrik Hildebrandt, Yagi Masafumi, Samuel Schmidt, and Ichi Tanaka for helpful discussions. Also, Adam Tomczak for providing the PSF matching code. 
C.E.R and C.D.F. were funded through the NSF grant AST-1312329, ``Collaborative Research: Accurate
cosmology with strong gravitational lens time delays'', and the HST grant
GO-12889. 
DS acknowledges funding support from a {\it {Back to Belgium}} grant from the Belgian Federal Science Policy (BELSPO). 
S.H. acknowledges support by the DFG cluster of excellence \lq{}Origin and Structure of the Universe\rq{} (\href{http://www.universe-cluster.de}{\texttt{www.universe-cluster.de}}).
 K.C.W. is supported by an EACOA Fellowship awarded by the East Asia Core Observatories Association, which consists of the Academia Sinica Institute of Astronomy and Astrophysics, the National Astronomical Observatory of Japan, the National Astronomical Observatories of the Chinese Academy of Sciences, and the Korea Astronomy and Space Science Institute. 
 S.H.S. acknowledges support from the Max Planck Society through the Max Planck Research Group. This work is supported in part by the Ministry of Science and Technology in Taiwan via grant MOST-103-2112-M-001-003-MY3.
 T.T. thanks the Packard Foundation for generous support through a Packard Research Fellowship, the NSF for funding through NSF grant AST-1450141, ``Collaborative Research: Accurate cosmology with strong gravitational lens time delays''. 
 LVEK is supported in part through an NWO-VICI career grant (project number 639.043.308).

Data analysis was in part carried out on common use data analysis computer system at the Astronomy Data Center, ADC, of the National Astronomical Observatory of Japan, as well as the SLAC National Accelerator Laboratory.

This work is based in part on observations obtained with MegaPrime/MegaCam, a joint project of CFHT and CEA/IRFU, at the CFHT which is operated by the National Research Council (NRC) of Canada, the Institut National des Sciences de l'Univers of the Centre National de la Recherche Scientifique (CNRS) of France, and the University of Hawaii. 
It is also based in part on observations made with the Spitzer Space Telescope, which is operated by the Jet Propulsion Laboratory, California Institute of Technology under a contract with NASA, and on observations obtained at the Gemini Observatory, which is operated by the Association of 
Universities for Research in Astronomy, Inc., under a cooperative agreement with 
the NSF on behalf of the Gemini partnership: the National Science Foundation 
(United States), the National Research Council (Canada), CONICYT (Chile), the 
Australian Research Council (Australia), 
Minist\'{e}rio da Ci\^encia, Tecnologia e 
Inova\c c\~ao (Brazil) 
and Ministerio de Ciencia, Tecnolog\'{i}a e Innovaci\'{o}n Productiva 
(Argentina).

The authors recognize and acknowledge the very significant cultural role and reverence that the summit of Mauna Kea has always had within the indigenous Hawaiian community. We are most fortunate to have the opportunity to conduct observations from this superb mountain.

TOPCAT \citep{taylor05} was used for catalogue matching. The codes developed during the course of this work are publicly available at \url{https://github.com/eduardrusu/zMstarPDF}.

\appendix

\section{Exploring systematics and sources of noise in the estimation of weighted count ratios}\label{section:systematics}

When measuring the weighted count ratios as described in Section \ref{section:ratiosdescript}, we account for several factors and estimate how much they contribute to the total uncertainty:

\begin{figure}
\includegraphics[width=90mm]{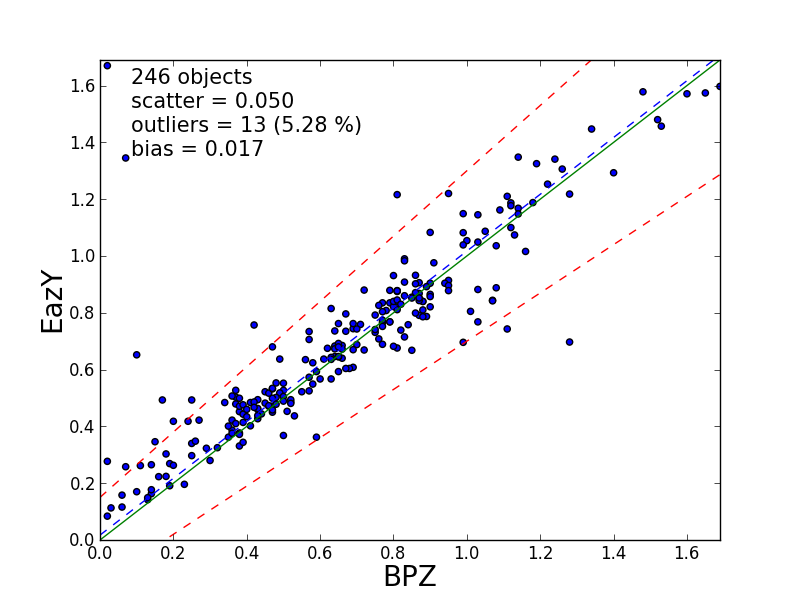}
\caption{Comparison of photometric redshift estimated with BPZ and EAZY  from $ugriJHK_s$ photometry, for the $i \leq 24$ galaxies within $120\arcsec$, without available spectroscopic redshifts.}
\label{fig:photoz}
\end{figure}

\begin{figure*}
\includegraphics[width=170mm]{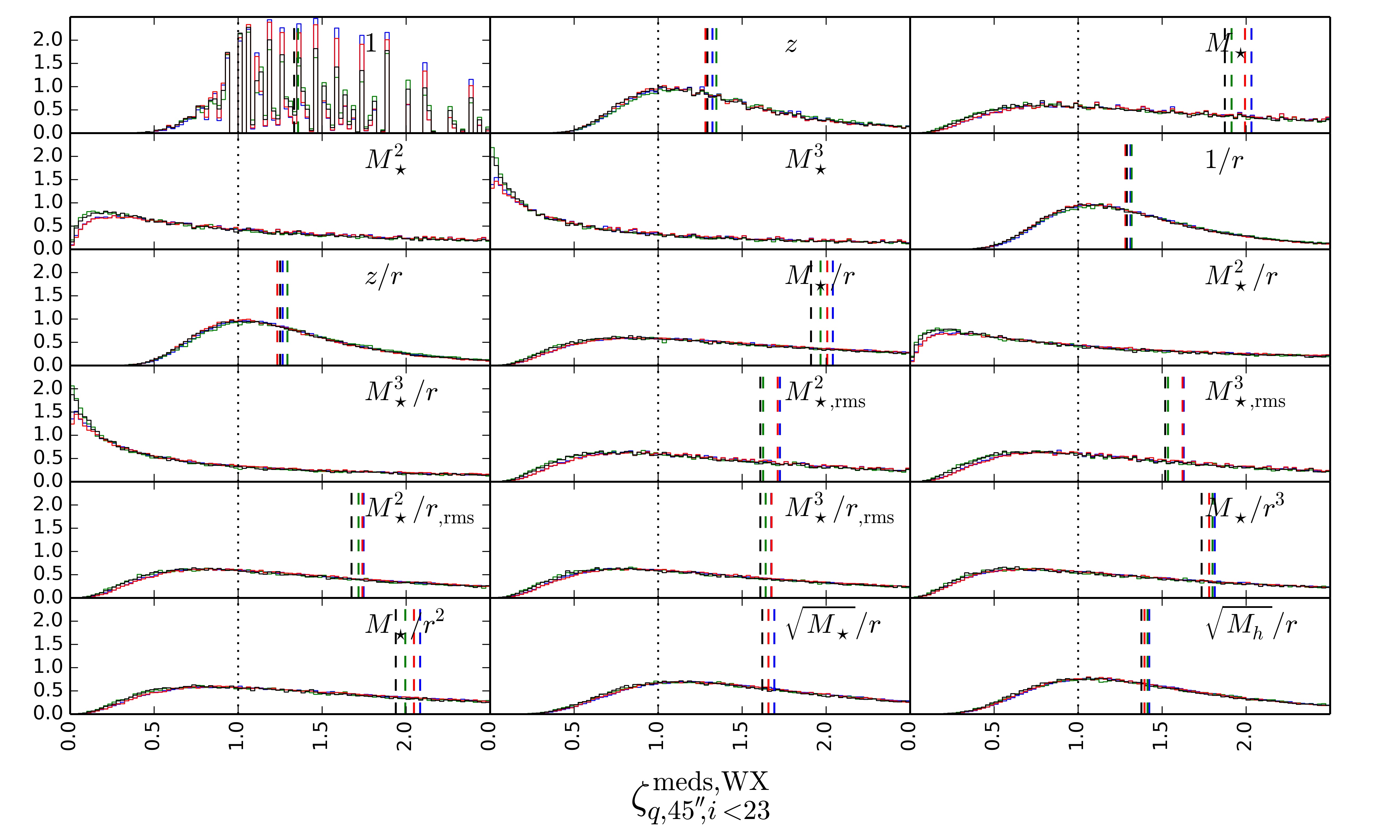}
\caption{Histograms of weighted count ratios for all $\zeta_q^\mathrm{meds,WX}$ weights, for galaxies inside a $45\arcsec$-radius aperture and $i\leq 23$. We use the plotting range and colors from Figure \ref{fig:12024meds}. The $q=1$ distribution appears discrete because of the small range of (positive integer) galaxy counts inside this small aperture and bright magnitude limit.
\label{fig:4523meds}}
\end{figure*}

\begin{figure*}
\includegraphics[width=170mm]{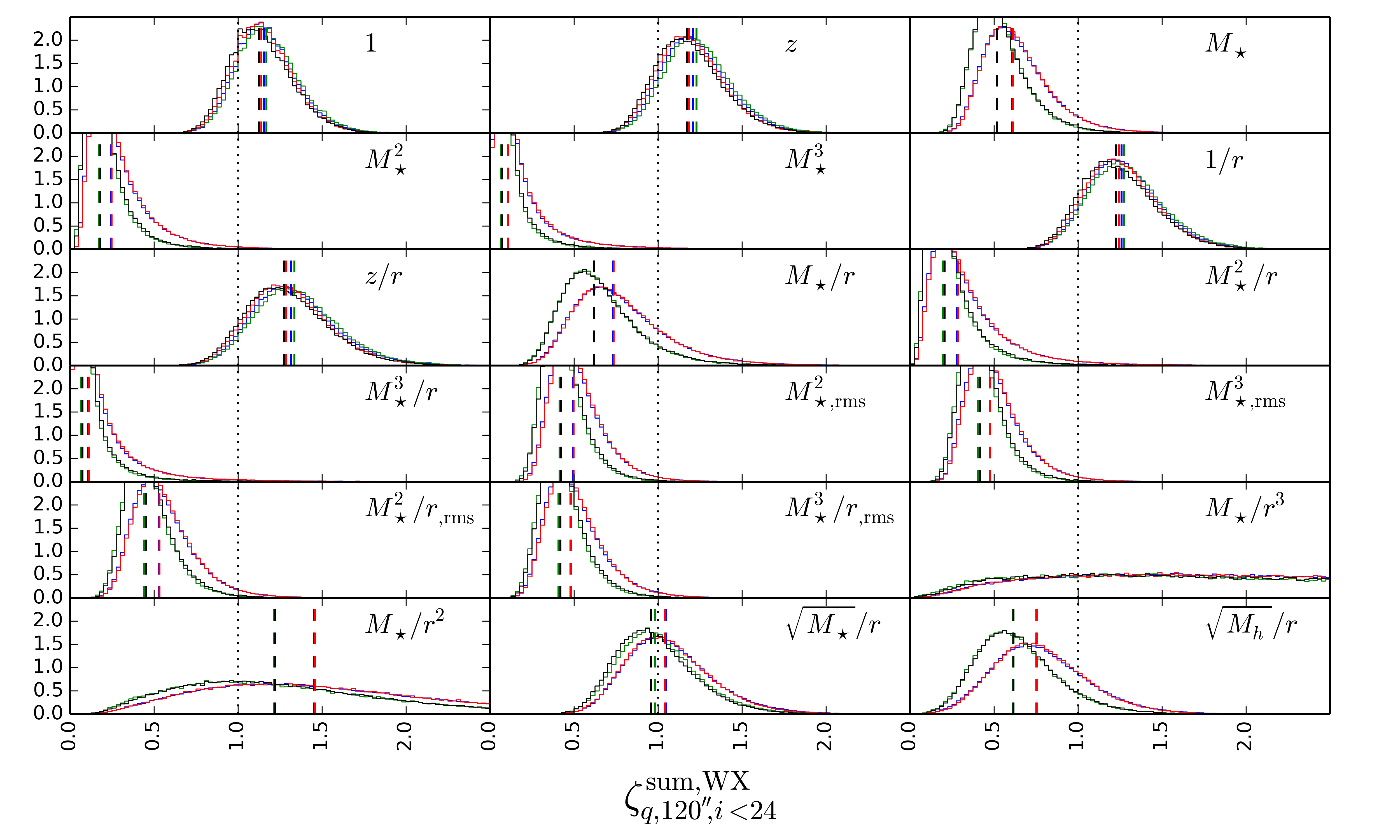}
\caption{Histograms of weighted count ratios for all $\zeta_q^\mathrm{sum,WX}$ weights, for galaxies inside a $120\arcsec$-radius aperture and $i\leq 24$. We use the plotting range and colors from Figure \ref{fig:12024meds}.
\label{fig:12024sum}}
\end{figure*}

\begin{itemize}
\renewcommand{\labelitemi}{$\bullet$}
\item Sample variance. To test the extent to which we are affected by sample variance, as well as by different fractions of stars in the CFHTLenS fields, we do not combine the W1-W4 fields, but measure overdensities for each of them separately. W4 is known to contain a larger fraction of stars \citep[][]{hildebrandt12}, and this may impact our results, given our galaxy-star classification, which assumes that all faint objects are galaxies (Section \ref{section:CFHT}). We also expect this to be the case for W2, given its low galactic latitude.\footnotemark \footnotetext{We used the plots available at \url{http://www.iac.es/proyecto/frida/skyCoverage.html} to estimate the relative number of stars, given the galactic coordinates of each field.} 
\item Fraction of masks. Using CFHTLenS cells with a substantial fraction of their areas covered by masks may introduce large Poisson noise. To estimate this effect, we exclude all cells that have more than  25\% and 50\%, respectively, of their areas masked. This results in eliminating $40\%, 32\%$ of the cells ($45\arcsec$ apertures), and $36\%, 24\%$ of the cells ($120\arcsec$ apertures), respectively.
\item Limiting magnitude and aperture radius. To quantify the dependence of our results on the aperture radius and limiting magnitude, we also consider limits of $45\arcsec$-radius (used by G13) and $i\leq23$ mag (S/N $\sim 30$), in addition to $120\arcsec$ and $i\leq24$.
\item Detection efficiency. In order to avoid biases when estimating weighted counts relative to CFHTLenS, and in view of the similarity between our $i$-band data for \hequad\ and the CFHTLenS $i$-band, we used the same detection parameters employed for the latter. However, galaxy counts at the limiting magnitude are sensitive to the detection parameters, and we found that by changing the DETECT\_THRESHOLD parameter in SExtractor from 1.5 to 2.5, we obtain more robust detections. We therefore consider the scatter between the two detection runs, where for each one we compute weighted ratios for all weights. 
\item Detections at the limiting magnitude. Due to uncertainties in the photometry at the limiting magnitude, some galaxies above the magnitude cut are in fact wrongly included in the cut, and vice-versa. This may bias the results. Therefore for all galaxies in the \hequad\ field we consider a gaussian around their SExtractor-measured $i$-band magnitude, with a standard deviation equal to the size of the photometric error bar, and randomly sample from this to test if the galaxy survives the color cut. We do this for each $i$ cell in a CHFTLenS WX field, as we compute $\zeta_q^\mathrm{WX} \equiv \left\{ W_q^{\mathrm{lens, mask}_\mathrm{i}}/W_q^{i\in\mathrm{WX}} \right\}$. It is unnecessary to do the same for the galaxies inside CFHTLenS, due to the large number of cells. 
\item Cell number dependence on the aperture radius. When considering a larger aperture radius around the lens system, and therefore a larger cell size, there are comparatively fewer contiguous non-overlapping cells spanning CFHTLenS. As a result, the $\zeta_q$ distribution will look noisy. To avoid this, we allow cells to partially overlap, with larger overlapping fraction for larger apertures. In practice, we use 2 equally spaced overlaps along each dimension of the $45\arcsec$-length cells (i.e. along each dimension in the grid, we consider cells centered at length/2, $2\times$length/2, $3\times$length/2 etc.), and 5 overlaps for the $120\arcsec$-length cells, respectively.
\item Different photometric redshift codes, and the importance of the IRAC bands. We include the scatter in the overdensities measured when using BPZ and EAZY separately, to compute photometric redshifts. This potentially affects more than just the weights explicitly incorporating redshift, since we do a cut at the source redshift, and the redshift values also affect the goodness-of-fit used to separate stars from galaxies. We also compute weights for stellar masses calculated with the inclusion of the IRAC channels, as well as without.
\item Accounting for the $P(z)$ and $P(M_\star | z)$ of an individual galaxy. Instead of just using the best-fit photometric redshift and median stellar mass for each galaxy in the \hequad\ field, we sample 10 times from the galaxy's redshift probability distribution, and compute the associated stellar mass (for which we also sample from the distribution returned by Le PHARE). We then compute $\zeta_q^{WX}$ for each of these. Again, it is not necessary to do this for the galaxies inside CFHTLenS, due to the large number of cells, which are only used once. 
\end{itemize}

\section{Details on inferring weighted count ratios from the MS}\label{section:MSdetails}

Even though we have made every effort to analyze the simulated data in
the same manner as the real data, this was not always possible, due to
inherent differences and computational reasons.  Here we present
details of our weighted count ratios estimation from the MS, and the
way the approach differs from the real data.

\begin{itemize}
\renewcommand{\labelitemi}{$\bullet$}
\item The MS catalogues represent a pure and complete sample of galaxies, whereas this is not the case in the real data. As a result, we randomly inject stars and remove galaxies, mirroring the contamination and incompleteness found in the real data. For this, we use the contamination and incompleteness fractions estimated in Figure 9 of \citet{hildebrandt12} for the CFHTLenS W1 field, as a function of magnitude. We considered 500 real stars for each 0.5 mag bin from CFHTLenS, and computed for these ``redshifts'' with BPZ, as well as ``stellar masses'' with Le PHARE. We then selected from these based on the the contamination fraction, and inserted them at random positions into each aperture of the simulation. 
\item It is important to use all the complete spatial extent of the MS (i.e. all MS fields), as our use of multiple conjoined weights when selecting lines of sight of similar overdensities (which we describe in Section \ref{section:kappa}) implies that we are limited by the number of available $\kext$ points found in the simulation. Each of the 64 MS fields has a corresponding $4096 \times 4096$ grid of convergence values (we refer to these as $\kext$ points), with $\sim 3.5\arcsec$ spacing. In Section \ref{section:systematics} we described how we use overlapping cells across the CFHTLenS fields. Here we use even higher fractions of overlaps, as we center one cell on each of the $\kext$ points. The only exceptions are at the edges of the fields, where the apertures would fall outside the field. 
\item Given the $\sim 10^9\ \kext$ points in the simulation, it is computationally expensive to estimate the weighted count ratios of each of the $45\arcsec$ or $120\arcsec$ aperture cells relative to every other cell, and take the median. 
In addition, the MS fields do not contain masks, in contrast to the \hequad \ and CFHTLenS fields. However, as we have seen in Section \ref{section:overdensities}, where we compared results after eliminating fields with different fractions of masks, the effect is negligible. The only masks we employ are the $5\arcsec$ radius inner masks around the center of each cell (to account for the fact that in the real data we masked the \hequad \ system itself, and its most nearby perturber), and the outer $45\arcsec$ or $120\arcsec$ radius representing the circular apertures. As a result, we can make an approximation in computing weighted count ratios. We compute the overdensity for each cell $i$ simply as $\zeta^{i,\mathrm{MS}}_q \equiv W^i_q/\overline{W^{i\in\mathrm{MS}}_q}$, where $\overline{W^{i\in\mathrm{MS}}_q}=\mathrm{median}(W^{i\in\mathrm{MS}}_q)$. We have checked that this redefinition is numerically indistinguishable from the one in Section \ref{section:ratiosdescript}, given the range spanned by $W_q$. We note, however, that the same approximation would not hold if we used the mean instead of the median. 
\item Due to the $JK_s$ field of view being slightly smaller than the $120\arcsec$ aperture, $\sim15\%$ of the galaxies around the edge of the \hequad \ field do not have coverage in these bands. We neglect this in the simulations. 
\item The MS catalogues do not contain synthetic magnitudes in the IRAC bands. However, as discussed in Section \ref{section:overdensities}, the effect that the exclusion of these bands has for the computation of weighting count ratios incorporating stellar masses is negligible.
\item Since we have a large number of cells, it is unnecessary to repeatedly sample from the magnitudes of the galaxies at the faint limit, like we did for the \hequad \ field. We also do not sample from the $P(z)$ and $P(M_\star)$ of each galaxy in the \hequad -like fields. Finally, we limit ourselves to the use of BPZ for estimating photometric redshifts.
\end{itemize}

\section{Cosmology dependence of the external convergence estimates}\label{section:cosm_depend}

Using a simple galaxy bias model, we can obtain a rough estimate of the cosmology dependence of the external convergence inferred from weighted galaxy counts. To first order in matter density fluctuations, the convergence $\kappa(\vtheta, \zs)$ for sources with angular image position $\vtheta$ and redshift $\zs$ can be expressed by a weighted projection of the matter density contrast $\deltam$ along the line of sight:
\begin{equation}
\label{eq:convergence}
\begin{split}
  \kappa(\vtheta, \zs) &=
  \frac{3 H_0^2\Om}{2 \mathrm{c}^2} 	
  \int_{0}^{\chis} \!\!\mathrm{d}\chid
  \,  \bigl(1 + \zd \bigr)
  \frac{\cadds \cadd}{\cads}
	\\&\quad\times
 \deltam \big(\cadd \vtheta, \chid, \zd \big)
  \\&=
 \frac{3 \Om}{2} 	
  \int_{0}^{\zs} \!\!\mathrm{d}{\zd}
  \,  \bigl(1 + \zd \bigr) \,
	\frac{H_0}{H(\zd)}
  \frac{\cadds \cadd}{\cads \chi_{H_0}}
		\\&\quad\times
 \deltam \big(\cadd \vtheta, \chid, \zd \big)	
	.
	\end{split}
\end{equation}
Here, $\mathrm{c}$ denotes the speed of light, $\chid=\chi(\zd)$, $\chis=\chi(\zs)$, $\cadd=\cadK(\chid)$, $\cads=\cadK(\chis)$, and $\cadds=\cadK(\chis-\chid)$, where $\chi(z)$ denotes the comoving line-of-sight distance for sources at redshift $z$, and $\cadK(\chi)$ the comoving angular diameter distance for comoving line-of-sight distance $\chi$. Furthermore, $\deltam(\vx, \chi, z)$ denotes the matter density contrast at comoving transverse position $\vx$, comoving line-of-sight distance $\chi$, and cosmic epoch expressed by the redshift $z$. 
Moreover, $\chi_{H_0} = \mathrm{c}/{H_0}$ denotes the Hubble distance, and $H(z)$ denotes the Hubble parameter at redshift $z$.

In a simple linear deterministic galaxy model, the galaxy density contrast $\deltag$ is related to the matter density contrast $\deltam$ by the relation:
\begin{equation}
	 \deltag \big(\cadd \vtheta, \chid, \zd \big)	 = \biasg \deltam \big(\cadd \vtheta, \chid, \zd \big)	
\end{equation}
with the galaxy bias parameter $\biasg$ as proportionality factor (assumed independent of redshift for simplicity).
Assume that the large-scale galaxy correlations and/or power spectra have been observed and their amplitude has been quantified, e.g. by a galaxy fluctuation amplitude parameter $\sigmag$ defined as the standard deviation of the galaxy density contrast $\deltam$ averaged over spheres of $8\,\mathrm{Mpc}$. The analogous quantity for the matter density contrast is the cosmic matter fluctuation amplitude $\sigma_8$. For a given cosmological model and observed galaxy clustering amplitude, the bias parameter can thus be expressed as $\biasg ={\sigmag}/{\sigma_8}$.
Hence,
\begin{equation}
\begin{split}
  \kappa(\vtheta, \zs) &=
 \frac{3 \Om \sigma_8}{2 \sigmag} 	
  \int_{0}^{\zs} \!\!\mathrm{d}{\zd}
  \,  \bigl(1 + \zd \bigr) \,
	\frac{H_0}{H(\zd)}
  \frac{\cadds \cadd}{\cads \chi_{H_0}}
		\\&\quad\times
 \deltag \big(\cadd \vtheta, \chid, \zd \big)	
	.
	\end{split}
\end{equation}

For all considered cosmologies, $\cads$, $\cadd$, and $\cadds$ are proportional to $H_0^{-1}$, and weakly varying with the cosmic mean density parameters $\Om$, $\OL$, etc. and equation-of-state parameters $w$, $w_0$, or $w_a$. Furthermore, $H(z)\propto H_0$ and weakly varying with $\Om$, $\OL$, $w$, etc. Thus, to lowest order in cosmological parameters, the convergence inferred from an observed galaxy density contrast $\deltag$ (or similar relative galaxy density quantities such as the weighted counts $\zeta_q$ considered in this paper) can be expressed by:
\begin{equation}
	  \kappa(\vtheta, \zs) = \frac{\Om}{\Om^{(0)}} \frac{\sigma_8}{\sigma_8^{(0)}} \kappa^{(0)}(\vtheta, \zs),
\end{equation}
where $\kappa^{(0)}(\vtheta, \zs)$ denotes the inferred convergence assuming cosmological parameters $\Om^{(0)}$ and $\sigma_8^{(0)}$ instead of parameters $\Om$ and $\sigma_8$, resp. Therefore, for an arbitrary function $F$ which depends on the external convergence $\kext$, this implies:
\begin{equation}
	\int \!\!\mathrm{d}\kext F(\kext, \ldots)
	=
		\int \!\!\mathrm{d}\kext^{(0)} F(\frac{\Om}{\Om^{(0)}} \frac{\sigma_8}{\sigma_8^{(0)}} \kext^{(0)}, \ldots)
,
\end{equation}
where $\kext^{(0)}$ denotes the external convergence inferred assuming $\Om^{(0)}$ and $\sigma_8^{(0)}$.

\bsp

\label{lastpage}
\end{document}